\documentclass[
reprint,onecolumn,
superscriptaddress,
amsmath,amssymb,
aps
]{revtex4-2}

\usepackage{graphicx}    
\usepackage{dcolumn}    
\usepackage{bm}         
\usepackage{float}       
\usepackage{subfigure}   
\usepackage{booktabs}   
\usepackage{multirow}   
\usepackage{tabularx}  
\usepackage{mathtools}  
\usepackage{physics}    
\usepackage{microtype} 
\usepackage{xcolor}     
\usepackage{siunitx}    
\usepackage{hyperref}  
\hypersetup{
  colorlinks=true,
  linkcolor=blue,
  citecolor=blue,
  urlcolor=blue
}

\usepackage{comment}

\begin{document}

\title{Dislocation dynamics on deformable surfaces}

\author{Marcello De Donno}
\affiliation{Institute of Scientific Computing, TU Dresden, 01062 Dresden, Germany}

\author{Luiza Angheluta}
\affiliation{Njord Centre, Department of Physics, University of Oslo, 0371 Oslo, Norway}

\author{Marco Salvalaglio}
\email{marco.salvalaglio@tu-dresden.de}
\affiliation{Institute of Scientific Computing, TU Dresden, 01062 Dresden, Germany}
\affiliation{Dresden Center for Computational Materials Science (DCMS), TU Dresden, 01062 Dresden, Germany}

\begin{abstract}
We develop a fully coupled theoretical description of dislocation dynamics on deformable crystalline surfaces, using continuum modeling and the amplitude-phase-field crystal (APFC) framework extended to curved geometries. We derive a general kinematic expression for dislocation velocity directly from the complex-amplitude evolution equations, which is also applicable to deformed surfaces through curvature-modified differential operators. From numerical simulations, we show that even small out-of-plane deformations reshape the phenomenology of defect motion through curvature-induced self-propulsion, modified glide directions, and non-classical defect-defect interactions. Our results show how surface geometry profoundly influences defect dynamics and establish the surface-APFC model as a powerful framework for predicting and interpreting curvature-defect coupling across a wide range of systems, from stiff but deformable layers to soft matter surfaces and membranes that retain crystalline order.
\end{abstract}


\maketitle

\section{Introduction}

Topological defects, such as dislocations and disclinations, play a central role in the mechanical response, structural organization, and dynamical behavior of two-dimensional (2D) crystalline materials. Their influence depends sensitively on the rigidity of the underlying surface and, in particular, on whether the surface can deform out of plane. In crystalline layers with high bending rigidity, such as atomically thin elastic membranes supported on solid substrates, the curvature is effectively fixed or slowly varying, and the surface may be treated as rigid \cite{nelson2002defects,Seung1988}. For such hard crystalline surfaces, defect energetics and interactions are well described by continuum elasticity on a prescribed geometry.
Extensions have been developed to account for continuum elasticity effectively coupled to surface deformation, incorporating additional geometrical effects in suspended layers
\cite{Witten2007,Guinea2008,Zhang2014a,Roychowdhury2018}.

By contrast, many soft matter systems, including colloidal monolayers on fluid interfaces, polymeric or lipid membranes, and biological tissues, possess low bending rigidity, so that surface geometry becomes an essential dynamic degree of freedom rather than an externally imposed constraint or secondary effect. Elasticity and geometry are dynamically coupled, such that soft crystalline surfaces can wrinkle, buckle, or form pleats to relax in-plane stresses, and the energetics and mobility of defects become strongly curvature-dependent. Local curvature redistributes stresses around defects and, in turn, can be actively generated by defect motion. Such geometry-defect coupling is particularly prominent in colloidal crystals with spatially varying curvature \cite{irvine2010pleats}, during tissue development where topological defects regulate morphogenetic events \cite{saw2017topological,angheluta2025topological}, and in actively driven elastic sheets, where defect-mediated stresses can reshape the substrate \cite{ho2024role}. These observations illustrate that the traditional paradigm of treating defects on flat, rigid surfaces is inadequate for soft crystalline materials.

This growing body of experimental and numerical evidence for soft surfaces and membranes retaining crystalline order underscores the need for theoretical frameworks that can simultaneously capture crystalline order, topological defects, and the evolving geometry of deformable surfaces. At large scales, defects in flexible crystalline membranes have been studied using generalized continuum elasticity theories \cite{nelson2002defects,bowick2009two}, including formulations based on the Föppl-von Kármán (FvK) equations that couple in-plane strain to bending-induced stretching \cite{Amar1997,nelson2002defects}. Although successful in predicting stress fields on curved surfaces, such continuum descriptions treat dislocations only indirectly through singular strain sources and therefore cannot resolve defect cores, dynamically evolving defect-defect interactions, or the feedback between defect activity and surface shape. As a consequence, they become insufficient in regimes where curvature and defects coevolve, as in soft membranes, thin films, and active elastic sheets.

Mesoscale modeling approaches address these limitations by directly resolving crystalline order. The phase-field crystal (PFC) model \cite{Elder2002,Elder2004,Emmerich2012} and its coarse-grained formulation through amplitude-expansion (APFC) \cite{Goldenfeld2005,salvalaglio2022coarse} provide an efficient description of slowly-varying lattice distortions and naturally represent dislocations as topological defects in the complex amplitude fields, without resolving atomic-scale features. Amplitudes themselves also serve as prominent descriptors for studying deformation and, importantly, defects in smooth field theories  \cite{Mazenko97,SkogvollNPJ2023,DeDonno2024}. Recent extensions of the PFC model have been proposed to describe crystals on curved and deformable surfaces \cite{Backofen2011,Kohler2016,Elder2021}, introducing curvature-modified differential operators and bending energies and enabling a fully coupled treatment of crystalline order and evolving surface geometry. This approach has also recently been realized in the coarse-grained APFC framework \cite{Benoit-Marechal_SurfaceAPFC}, accessing continuum length scales while retaining key details of crystal lattices and their defects.

In this work, we exploit descriptions based on classical continuum mechanics combined with recent developments in PFC modeling to investigate the dynamics of dislocations on deformable crystalline surfaces. 
In particular, we derive a general kinematic framework for surface deformation based on an energy-driven continuum description in a surface-height formulation, employing APFC modeling to capture mesoscale elasticity and defect dynamics. 
This formulation enables quantitative characterization of curvature-induced self-propulsion of defects, deviations from classical glide trajectories, and the interplay between bending rigidity, in-plane stresses, and defect topology. Through numerical simulations, we show that even basic out-of-plane deformations can qualitatively alter dislocation motion, underscoring the importance of fully coupled surface-defect descriptions in soft crystalline surfaces. 

The paper is organized as follows. In Section~\ref{sec:continuum_modeling}, we review the continuum mechanics of deformable crystalline surfaces within the FvK framework. We then introduce the APFC modeling framework that couples crystalline order with surface deformations in Section~\ref{sec:apfc}. In Section~\ref{sec:velocity_derivation}, we analytically derive an expression for dislocation velocities in terms of the amplitude dynamics. Numerical results are then discussed in Section~\ref{sec:results_selfpropulsion} concerning self-propulsion of dislocations emerging from the coupling of surface deformation and in Section~\ref{sec:results_surfaces} concerning general traits of dislocation dynamics on fixed curved profiles. A semi-analytical formulation and analysis exploiting controlled approximations is reported in Section~\ref{sec:analytica}. We conclude with a summary and discussion in Section~\ref{sec:conclusion}.

%-------------
\section{Continuum modeling of deformable surfaces}
\label{sec:continuum_modeling}

We consider a deformable surface (or membrane) $\mathcal{S}$ that exhibits crystalline order and is free to undergo out-of-plane deformations. Its geometry is tracked as a smooth height profile $h\equiv h(\mathbf{r})$ with $\mathbf{r} \equiv (x,y)\in \Omega \subset  \mathbb{R}^2$. Expressions of differential operators and norms evaluated on $\mathcal{S}\equiv(\mathbf{r},h(\mathbf{r}))$ are reported in the Appendix \ref{app:operators}.

The mechanical response of flexible surfaces arises from the coupled effects of in-plane stretching and out-of-plane bending. For thin plates, this balance can be captured by the energy functional of the Föppl-von Kármán (FvK) theory, $\mathcal F_{\rm FvK}$ \cite{Seung1988,Zhang2014a}
\begin{equation}
\begin{split}\label{eq:Ffvk}
\mathcal F_{\mathrm{FvK}} &= 
 \int_\Omega \mathrm d\Omega \, \frac{1}{2}\kappa(\nabla^2h)^2 
+ {\lambda} \delta_{ij}U_{kk}^2 + 2\mu U_{ij}^2,
\\
U_{ij} &= \underbrace{\frac{1}{2}\bigg(\partial_i u_j + \partial_j u_i\bigg)}_{\varepsilon_{ij}} + \frac{1}{2}(\partial_i h \partial_j h),
\end{split}    
\end{equation}
where $\kappa$ is the bending stiffness of the surface, $\mu,\,\lambda$ the Lamé parameters of the surface, and $\mathbf{u}\equiv(u_x,u_y)$ the displacement field defined in $\Omega$. 
Here, $\boldsymbol{\varepsilon}$ is referred to as the membrane (in-plane) strain, while the second term in $U_{ij}$ is the bending strain. Given this energy functional, the overdamped dynamics of the surface profile can be obtained by \cite{Nitschke2020,Benoit-Marechal_SurfaceAPFC}
%
%\begin{equation}\label{eq:dt_hei}
%\partial_t h = 
%-M_h \sqrt{|\mathbf{g}|} \frac{\delta \mathcal F}{\delta \xi},
%\end{equation}
%
\begin{equation}
\partial_t h =
-M_h \sqrt{|\mathbf{g}|} \frac{\delta F_{\rm FvK}}{\delta h} =
M_h \sqrt{|\mathbf{g}|} \bigg(\kappa\nabla^4h +
\partial_i(\widetilde{\sigma}_{ij}\partial_jh)\bigg),  
\end{equation}
with $M_h$ a mobility coefficient for the height profile, $\mathbf{g} = 1 + (\nabla h)^2$ the metric tensor (first fundamental form), describing how distances on $\mathcal{S}$  differ from distances in the underlying parameter space $\Omega$, and $\xi:\Omega\rightarrow \mathbb{R}$ the normal function measuring the distance from $h(\mathbf{r})$ along the normal of $\mathcal{S}$.

The stress tensor $\widetilde{\sigma}$ corresponds to the sum of the membrane (in-plane) and bending stresses:
\begin{equation}\label{eq:FvP_stress}
\begin{split}
\tilde{\sigma}_{ij} &= \sigma_{ij} + \sigma_{ij}^h,
\\
\sigma_{ij} &=\lambda \delta_{ij}\varepsilon_{kk}+2\mu \varepsilon_{ij},
\\
\sigma_{ij}^h &= \mu (\partial_i h)(\partial_j h) +
    \delta_{ij}\frac{\lambda}{2}\bigg((\partial_x h)^2+(\partial_y h)^2\bigg).
\end{split}    
\end{equation} 
To get general insights into the contributions governing the dynamic, we can rewrite the evolution of the height profile as
\begin{equation}\label{eq:dth_KSH}
\begin{split}
\partial_t h &= 
{M_h}{\sqrt{|\mathbf{g}|}} \bigg[\beta + \tau + \tau^h \bigg],
\\
\beta &= \kappa \nabla^4 h,
\\
\tau &= (\nabla\cdot \boldsymbol \sigma )\cdot\nabla h + \mathrm{Tr}[\boldsymbol \sigma\cdot\nabla\nabla h],
\\
\tau^h &= (\nabla\cdot \boldsymbol \sigma^h )\cdot\nabla h + \mathrm{Tr}[\boldsymbol\sigma^h\cdot\nabla\nabla h].    
\end{split}
\end{equation}
Here, $\beta$ incorporates the surface bending stiffness and depends solely on the height profiles. This term encodes the tendency of the film to flatten, i.e., to minimize surface curvature. $\tau$ couples local in-plane stress to surface deformations; note that the first term is zero at mechanical equilibrium in the absence of any external forces (a condition indeed reading $\nabla \cdot \boldsymbol \sigma=0$) while it mediates surface evolution when deviating from this limit. $\tau^h$ resembles the same form as $\tau$ but depends on the bending stress $\sigma^h$. 
In the considered parametrization, this latter term can then be evaluated further by expanding $\sigma^h$ in gradients of $h$ using the definition in Eq.~\eqref{eq:FvP_stress}:
\begin{equation}
\begin{split}
\tau^h = 
\bigg(\mu + \frac{\lambda}{2} \bigg) \bigg[
\bigg(3 (\partial_x h)^2 + (\partial_y h)^2\bigg) \partial_{xx} h
+ \bigg((\partial_x h)^2 +3 (\partial_y h)^2 \bigg) \partial_{yy} h 
+ 4 \, \partial_x h \, \partial_y h \, \partial_{xy} h 
\bigg].    
\end{split}
\end{equation}
Therefore, it results in a contribution that depends only on the variation in the height profile. As this preliminary analysis shows, the deformation field is strongly coupled with out-of-plane deformations. In this work, we aim to describe mesoscale scenarios in which we track isolated dislocations (different from considering dislocation through continuum descriptions \cite{singh2022interaction}) and the impact of out-of-plane deformations using a suitable model that encodes the FvK equations, as described in the following section.

\section{Amplitude-PFC modeling}\label{sec:apfc}

The PFC model enables the representation of crystal lattices over diffusive time scales \cite{Elder2002,Elder2004,Emmerich2012} by introducing an energy functional $\mathcal{F}_{\rm PFC}$ minimized by a periodic order-parameter field $\psi(\mathbf r,t)$, which encodes the atomic density, and prescribing its evolution through a conservative gradient-flow dynamics. This framework captures elastic effects in crystals that retain microscopic length scales, although the explicit reconstruction of elastic fields and tracking of dislocations generally requires numerical coarse-graining and post-processing. However, a formal coarse-graining can instead be achieved by focusing on the complex amplitudes of the principal Fourier modes and their dynamics, resulting in the so-called Amplitude PFC (APFC) model \cite{Goldenfeld2005,salvalaglio2022coarse}. This formulation accounts for elasticity and dislocation behavior in a self-consistent manner, as the amplitudes vary on scales compatible with the continuum description of elastic fields while simultaneously providing a natural representation of dislocations.

In brief, the atomic density $\psi(\mathbf r,t)$ is approximated as a superposition of plane waves with slowly varying complex amplitudes $\eta_n$
\begin{equation}\label{eq:amplitude_expansion}
    \psi(\mathbf r,t) = \sum_{n=1}^N \eta_n(\mathbf r,t)e^{i\mathbf{k}_n\cdot\mathbf{r}} + \mathrm{c.c.}, 
\end{equation}
where $\mathbf{k}_n$ are the $N$ principal reciprocal space vectors corresponding to a specific lattice symmetry and $\mathrm{c.c.}$ is the complex conjugate. Spatial and time dependence of amplitudes will be omitted hereafter for brevity. The free energy for APFC on a (flat) domain $\Omega$, which can be obtained by coarse-graining the PFC free energy, $F_{\rm PFC}[\psi]$, reads \cite{salvalaglio2022coarse}
\begin{equation}\label{eq:free-energy-flat}
    \mathcal{F}_{\rm APFC}[\{\eta_n,\eta_n^*\}] = \int_{\Omega} \bigg[A\sum_{n=1}^N |\mathcal{G}_n\eta_n|^2 + g(\{\eta_n,\eta_n^*\})\bigg]\ d\mathbf{r},
\end{equation}
where $A$ is a model parameter scaling the elastic constants, $g(\{\eta_n,\eta_n^*\})$ a degree-four polynomial of the amplitudes and their complex conjugates, and $\mathcal G_n = \nabla^2 + 2i\mathbf{k}_n\cdot\nabla$. 
In this work, we consider triangular lattices, for which $N=3$ and 
$\mathbf{k}_1 = (-\sqrt{3}/2,-1/2),\,
\mathbf{k}_2 = (0,1),\,
\mathbf{k}_3 = (\sqrt{3}/2,-1/2)$.
Then, in Eq.~\eqref{eq:free-energy-flat}, the polynomial in the amplitudes $g\{\eta_n,\eta_n^*\}$ reads
\begin{equation}
    g\{\eta_n,\eta_n^*\}^\mathrm{TRI} = 
    B_2 \Phi 
    + B_3 \bigg(\eta_1 \eta_2 \eta_3 + \eta_1^* \eta_2^* \eta_3^* \bigg) 
    + B_4 \bigg( 3\Phi^2 - 6 \sum_{n=1}^3 |\eta_n|^4 \bigg),
\end{equation}
with $B_2,\,B_3,\,B_4$ model parameters, and 
\begin{equation}\label{eq:amod}
    \Phi = \sum_{n=1}^3 |\eta_n|^2,
\end{equation}
an order parameter that is constant in bulk and decreases at dislocations. The evolution of amplitudes approximating the conserved dynamics of $\psi$ is given by \cite{salvalaglio2022coarse}
\begin{equation}
    \partial_t \eta_n = - \|\mathbf{k}_n\|^2 \frac{\delta F}{\delta \eta_n^*}. 
\end{equation}

We consider the extension of the APFC framework to describe deformable surfaces (sAPFC) introduced in Ref. \cite{Benoit-Marechal_SurfaceAPFC}. This approach exploits the height-formulation description outlined in Section~\ref{sec:continuum_modeling}, thereby evaluating differential operators and norms as in \eqref{eq:surf_op} and introducing an additional term that accounts for surface bending. The free energy functional reads
\begin{equation}\label{eq:free-energy-surf}
    \mathcal{F}_\mathcal{S}
    = \underbrace{\int_\mathcal{S} \bigg[A\sum_{n=1}^N |\mathcal{G}_n^\mathcal{S}\eta_n|^2\bigg]\ d\mathcal{S}}_{\mathcal{F}_{\rm elas}} + \int_\mathcal{S} \bigg[ g(\{\eta_n,\eta_n^*\})\bigg]\ d\mathcal{S} 
    + \underbrace{\int_\mathcal{S} \bigg[ \frac{\kappa}{2} \mathcal{H}^2 \bigg] d\mathcal{S}}_{\mathcal{F}_{\rm bend}},
\end{equation}
where the first two terms result from considering $\mathcal{F}_{\rm APFC}$ with surface operators, $\mathcal{F}_{\rm elas}$ encodes the elastic energy associated with in-plane deformation (as in the classical model formulation for flat spaces), $\mathcal{F}_\mathrm{bend}$ is the bending energy as in the Helfrich model \cite{Helfrich_bend}, with $\mathcal{H}$ the mean curvature of the surface and $\kappa$ the bending stiffness. This free energy reduces to \eqref{eq:Ffvk} for small deformations \cite{Benoit-Marechal_SurfaceAPFC}.
The operator $\mathcal G_n^S \eta_n$ reads
\begin{equation}
\mathcal G_n^S \eta_n  = \nabla^2_S \eta_n 
+ (\|\mathbf k_n\|^2-\|\mathbf k_n\|^2_S)\eta_n
+ i[2\langle \mathbf k_n,\nabla_S\eta_n \rangle_S
+ \mathrm{div}_S(\mathbf k_n)\eta_n]    .
\end{equation}
The dynamics of the amplitudes and the height profile are then given by
\begin{equation}\label{eq:surf_evo}
\begin{split}
\partial_t \eta_n &= (\partial_t h) \frac{\langle \nabla h, \nabla \eta_n \rangle}{|\mathbf{g}|}
- M_\eta 
\left( \|\mathbf{k}_n\|^2 - \frac{\langle \mathbf{k}_n, \nabla h \rangle^2}{|\mathbf{g}|} \right)
\frac{\delta \mathcal{F}_\mathrm{\mathcal{S}}}{\delta \eta_n^*},
\\
\partial_t h &= - M_h \sqrt{|\mathbf{g}|} 
\left( 
    \frac{\delta \mathcal{F}_\mathrm{elas}}{\delta \xi} + \frac{\delta \mathcal{F}_\mathrm{bend}}{\delta \xi} 
\right), 
\end{split}
\end{equation}
where $M_\eta$ is a mobility coefficient for the amplitudes.
Detailed derivation and discussion of this model can be found in Ref.~\cite{Benoit-Marechal_SurfaceAPFC}. 

% \textcolor{red}{MS: Two sentences on numerical simulations here}
The evolution equations are solved using a Fourier pseudo-spectral method~\cite{salvalaglio2022coarse,Benoit-Marechal_SurfaceAPFC}, allowing the linear operators to be treated exactly as algebraic functions of the wavevectors (coordinate of the Fourier space) while the nonlinear terms are computed from their Fourier transforms. 
This method imposes applying periodic boundary conditions.
Time integration is carried out using a first-order exponential time differencing scheme.
The code is implemented in Python and exploits the established Fast Fourier Transform library FFTW~\cite{FFTW}.

\section{Dislocation velocity from the evolution of amplitudes}
\label{sec:velocity_derivation}

A deformation of the crystal lattice by a displacement field $\mathbf{u}$ introduces a phase shift in the complex amplitudes entering \eqref{eq:amplitude_expansion}, which can be written as
\begin{equation}\label{eq:amplitude_displacement}
\eta_n=\phi_n e^{-i\mathbf{k}_n\cdot\mathbf{u}},
\end{equation}
i.e., with the phase $\theta_n=-\mathbf{k}_n\cdot\mathbf{u}$ and $\phi_n=|\eta_n|$ the amplitude magnitude. A dislocation with Burgers vector $\mathbf{b}$ is associated with a singular displacement field at the position of the core $\mathbf{r}_0$. This leads to a phase slip in the complex amplitude, corresponding to a topological charge $s_n$, given by 
\begin{equation}
    2\pi s_n=\oint_C d\theta_n = -\oint_C \mathbf{k}_n \cdot \mathrm{d}\mathbf{u}=-\mathbf{k}_n \cdot \mathbf{b},
\end{equation}
with $C$ a close circuit around $\mathbf{r}_0$ and $s_n=\pm1$ for perfect dislocations. In the APFC model outlined in Section~\ref{sec:apfc}, $\phi_n=0$ for $s_n \neq 0$, thus realizing an integrable free energy functional.

An amplitude $\eta(\mathbf r,t)$ carrying topological charge thus remains a single-valued complex field by having zero magnitude at the defect core $\mathbf r^0(t)$, where its phase becomes undefined, namely
\begin{equation}
    \eta\bigl(\mathbf r^0(t),t\bigr) = 0.    
\end{equation}
The conservation of topological charge means that an isolated defect is stable and it cannot vanish unless it interacts with another defect of opposite charge or is absorbed at a boundary~\cite{SkogvollNPJ2023}. Thus, in the comoving reference frame of the defect, its core must remain a zero of $\eta$ throughout the evolution. We can then write the following evolution law for the amplitude
\begin{equation}
    \frac{d}{dt}\,\eta\bigl(\mathbf r^0(t),t\bigr)
    =\partial_t\eta + \frac{d\mathbf r^0}{dt}\cdot\nabla\eta|_{\mathbf r=\mathbf r^0}
    =0.    
\end{equation}
with $\mathbf v = (v_x,v_y) = d\mathbf r^0/dt$ the the dislocation (core) velocity. In 2D, we get
\begin{equation}\label{eq:vel1}
    \partial_t \eta + v_x\,\partial_x\eta + v_y\,\partial_y\eta =0,    
\end{equation}
Solving this for $v_x,v_y$ for a given $\partial_t \eta$ allows us to determine the dynamics of the defect from the evolution of the order parameter. To this goal, we can solve the two equations obtained by first multiplying Eq.~\eqref{eq:vel1} by $\partial_x\eta^*$ and $\partial_y\eta^*$ and then taking the imaginary part. This results in 
%
% \begin{equation}
%     \mathrm{Im}\Bigl[
% \partial_y\eta^*\,(\partial_t\eta + v_x\,\partial_x\eta + v_y\,\partial_y\eta)
% \Bigr] = 0.    
% \end{equation}
% %
% Distribute:
% %
% \begin{equation}
% \mathrm{Im}
% \Bigl[\partial_y\eta^*\partial_t\eta \Bigr]
% + v_x\mathrm{Im}
% \Bigl[\partial_y\eta^* \partial_x\eta\Bigr] 
% + v_y\mathrm{Im}
% \Bigl[\partial_y\eta^* \partial_y\eta\Bigr] 
% = 0.    
% \end{equation}
%

%
%\begin{equation}
%\begin{split}
%    0 &= \mathrm{Im}\Bigl[ \partial_y\eta^*\,(\partial_t\eta + v_x\,\partial_x\eta + v_y\,\partial_y\eta)\Bigr]
%    \\[1ex]
%    &= \mathrm{Im}\Bigl[\partial_y\eta^*\partial_t\eta \Bigr]
%        + v_x\mathrm{Im}\Bigl[\partial_y\eta^* \partial_x\eta\Bigr] 
%        + v_y\mathrm{Im}\Bigl[\partial_y\eta^* \partial_y\eta\Bigr].
%\end{split}    
%\end{equation}
%
%The last term is zero, since 
%$\mathrm{Im}(\partial_y\eta^*\,\partial_y\eta) = \mathrm{Im}(|\partial_y\eta|^2)=0$.
%Then, we get:
%
\begin{equation}\label{eq:vv}
\begin{split}
v_x 
&= -\mathrm{Im}
\Bigl[\partial_y\eta^*\partial_t\eta \Bigr] \Big/ \mathrm{Im}
\Bigl[\partial_x\eta\,\partial_y\eta^* \Bigr] ,     
\\
v_y 
&= -\mathrm{Im}
\Bigl[\partial_x\eta^*\partial_t\eta \Bigr] \Big/ \mathrm{Im}
\Bigl[\partial_x\eta^* \partial_y\eta\Bigr] .     
\end{split}
\end{equation}
%
% Given an expression for the gradient of the amplitudes and their evolution, Eqs.~\eqref{eq:vv} can then be exploited to determine the velocity of dislocations. 
% In the case considered in this work, $\partial_t \eta_n$ also contains the time evolution and gradients of the surface profile; see Eq.~\eqref{eq:surf_evo}. 
% Therefore, this description allows for inspection of the impact of surface deformation on dislocation dynamics, as further discussed in the following.
Given an expression for the gradient of the amplitudes and their evolution, Eqs.~\eqref{eq:vv} can be used to determine the velocity of dislocations by superposition of the velocities associated with moving zeros of the amplitudes.
Crucially, $\partial_t \eta$ fully incorporates both the time evolution and the gradients of the surface profile (see Eq.~\eqref{eq:surf_evo}), meaning that the effects of curvature are automatically accounted for.
As a result, the same equations can be applied to flat or curved, rigid or deformable surfaces, allowing a unified description of how surface geometry influences dislocation dynamics, as further discussed below.

\section{Dislocation self-propulsion on deformable surfaces}
\label{sec:results_selfpropulsion}

\begin{figure}
    \centering
    \includegraphics[width=\linewidth]{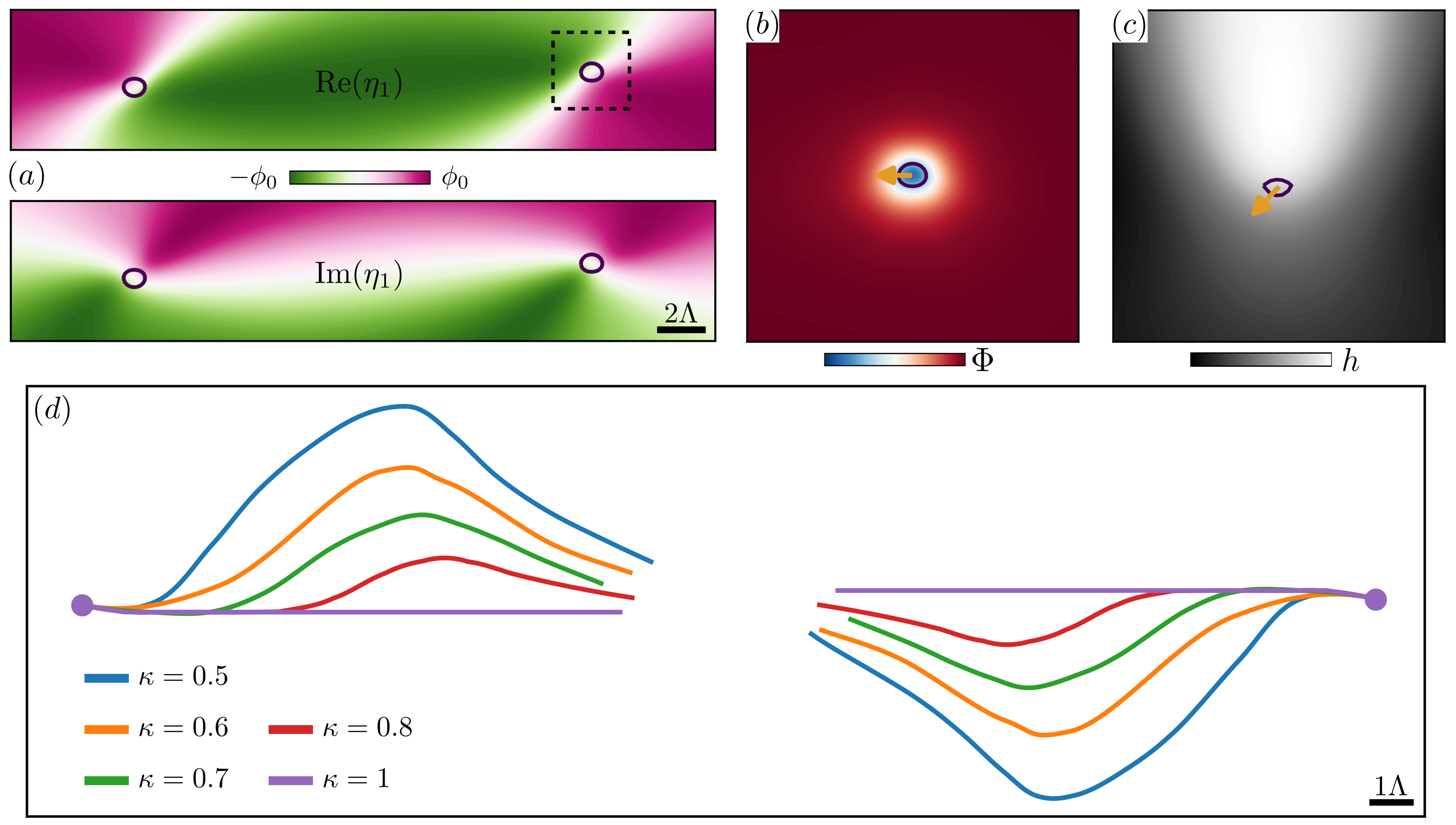}
    \caption{Dislocation dipole.
    (a): Real and imaginary parts of an amplitude carrying topological charge. 
    Isolines $\Phi = 0.3\,\Phi_\mathrm{MAX}$ enclose the dislocation cores. 
    $\phi_0$ is the (real) value of the amplitude in the relaxed bulk.
    (b): Field $\Phi$ in the square region highlighted in panel (a) by the dashed line.
    The black solid line corresponds to the isoline $\Phi = 0.3\,\Phi_\mathrm{MAX}$. 
    The orange arrow represents the velocity of the dislocation core calculated via Eq.~\eqref{eq:vv}.
    (c): Same setup as panel (b), but considering a deformable surface. The color map shows the height profile around the rightmost defect. 
    The black solid line corresponds to the isoline $\Phi = 0.3\,\Phi_\mathrm{MAX}$. 
    The orange arrow represents the velocity of the dislocation core calculated via Eq.~\eqref{eq:vv}. 
    (d): Trajectories of dislocation dipoles on deformable surfaces with variable bending stiffness $\kappa$.
    }
    \label{fig:velocity}
\end{figure}

We consider a triangular lattice with the reciprocal lattice vectors as modeled in Section~\ref{sec:apfc}, and a dislocation dipole with Burgers vectors $\mathbf{b} = \pm(4\pi/\sqrt{3}, 0)$, producing topological charges $s_1 = \mp 1$ and $s_3=\pm 1$. This can be initialized by exploiting Eq.~\eqref{eq:amplitude_displacement} with $\mathbf{u}$ the displacement field of a dislocation at equilibrium in an isotropic elastic body \cite{anderson2017,salvalaglio2022coarse} with corresponding $\mathbf{b}$ and position of the core. The dipole is positioned in a glide configuration, meaning that the Burgers vectors are chosen parallel to the glide direction along which dislocations move. The initial distance between the dislocations is 32 lattice parameters $\Lambda = \Lambda_x = 4\pi/\sqrt{3}$.

In the absence of out-of-plane deformation, the dislocations move horizontally (glide) one towards the other, until they annihilate.
A snapshot of the evolution is shown in Figure~\ref{fig:velocity}(a).
The position of each dislocation is represented by a black isoline of the order parameter $\Phi$ defined in Eq.~\eqref{eq:amod}. 
Figure~\ref{fig:velocity}(b) shows a close-up of the rightmost dislocation. The orange arrow indicates the velocity of the amplitude-zero, calculated according to Eq.~\eqref{eq:vv} with $\partial_t \eta$ evaluated from the simulation data as $\Delta\eta/\Delta t$, i.e., the variation of each amplitude over a simulation timestep.
Eq.~\eqref{eq:vv} thus consistently predicts the horizontal motion of dislocations along the glide plane as they approach one another, and the calculated velocities agree with those observed in the simulations.

We then repeat the simulation with the same setup on a surface that can bend (bending stiffness $\kappa = 0.5$). The defects still attract each other and eventually annihilate, but their trajectory now deviates from the glide plane. In Figure~\ref{fig:velocity}(c), we report a plot of the height profile produced by the rightmost dislocation.
The orange arrow shows the dislocation velocity calculated from Eq.~\eqref{eq:vv}. A pronounced vertical component emerges. This indicates that dislocations move in a direction perpendicular to the Burgers vector, driven by the surface bulge they create. 
A self-propulsion mechanism is realized through out-of-plane deformation at the core, which affects both the stationary states of isolated defects and the outcomes of their interactions (e.g., annihilation processes). 
The magnitude of this effect depends on the bending stiffness of the surface, which in turn determines the magnitude of the out-of-plane deformations. In Figure~\ref{fig:velocity}(d), we show the trajectories of dislocation dipoles evolving on surfaces with different bending stiffness. 
For large bending stiffness ($\kappa=1$), the trajectory is the same as for the flat case, owing to negligible out-of-plane deformations.
As the bending stiffness decreases, more and more significant bending is realized, and the dislocations move further away from the glide plane. 
This evidence matches previous qualitative observations \cite{Benoit-Marechal_SurfaceAPFC} and, importantly, it is quantitatively captured by Eq.~\eqref{eq:vv}. We finally note that the dislocations are located on the slope of the bulge they create. In Section~\ref{sec:analytica}, we propose a simplified semi-analytical description which produces the same effect, further supporting this evidence.

\section{Impact of surfaces with fixed profile}
\label{sec:results_surfaces}

To further examine how out-of-plane deformation affects dislocation motion, we now focus on a fixed surface profile and analyze its influence on dislocation velocities. 
We consider a triangular lattice $300\times100$ unit cells in size, where each unit cell has dimensions $\Lambda_x = 4 \pi /\sqrt{3}$, $\Lambda_y = 4 \pi$.
We set the height profile to a double Gaussian bump, defined as
\begin{equation}\label{eq:bumps}
  h(x,y) =a(e^{-((y-y_0)/\Delta)^2/2}+e^{-((y+y_0)/\Delta)^2/2}).  
\end{equation}
We take $y_0 = L_y/4$, with $L_y$ the domain size along the $y$-axis, so that the peak-to-peak distance of the Gaussians is $L_y/2 = 50 \Lambda_y$.
The peak height $a$ and the standard deviation $\Delta$ can be used to control the slope of the bumps. We set $a = 5 \Lambda_x$ and $\Delta=2\Lambda_y$. 
   
\begin{figure}
    \centering
    \includegraphics[width=\linewidth]{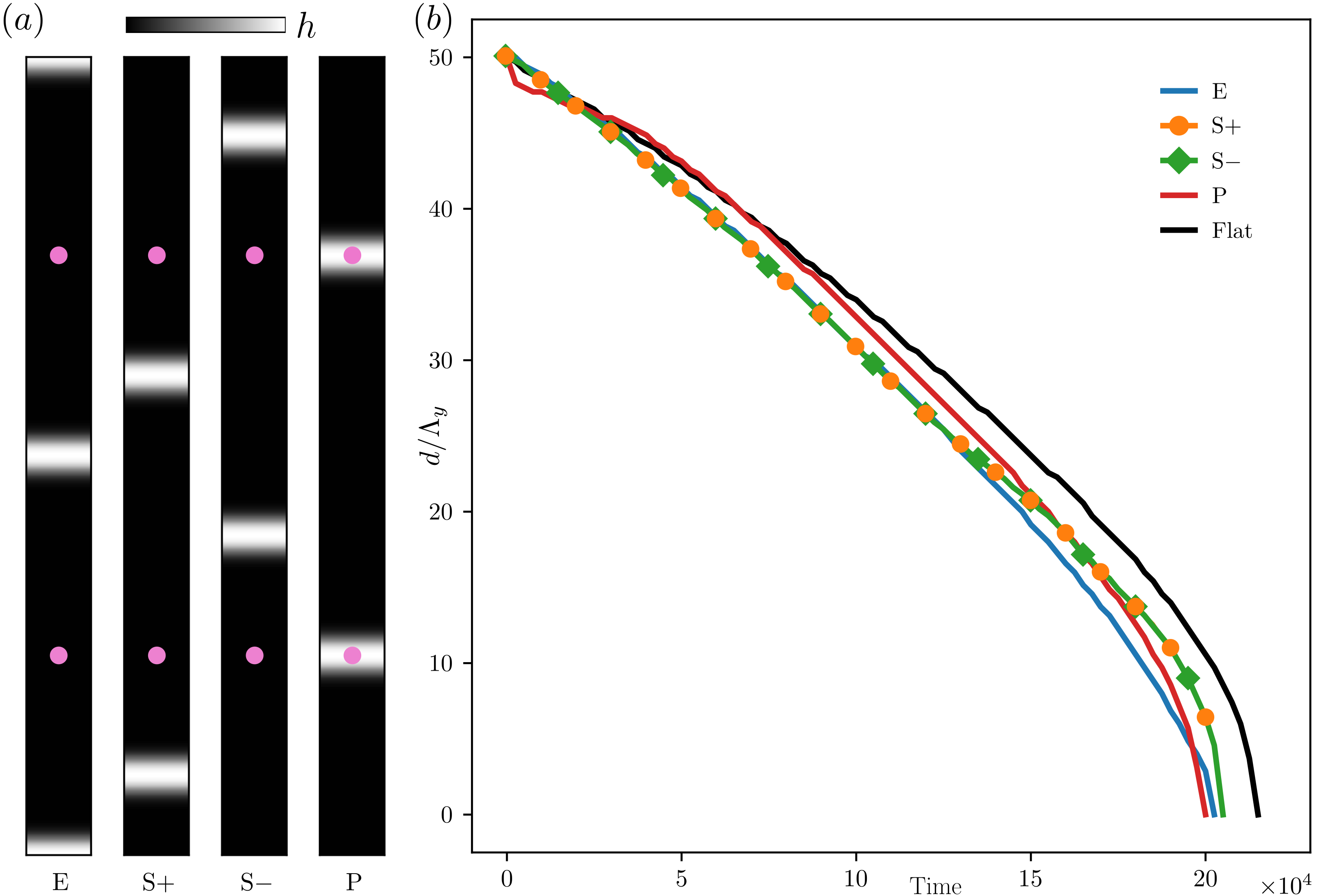}
    \caption{Effect of the curvature on dipole motion.
    (a) Illustration of the initial conditions. 
    The density plot shows the double-Gaussian height profile.
    Pink dots mark the initial positions of the dislocations.
    We define four configurations: dislocations starting from the flat, equidistant to the Gaussian profiles (E), starting from the flat, with one defect closer to the Gaussian than the other (S$+$, S$-$), and starting at the peak of the Gaussian profiles (P). 
    The initial distance between the dislocations is the same in all configurations. 
    (b) Distance between the dislocations over time in the four configurations illustrated in panel (a) as well as for a flat surface (Flat).
    %All systems on a curved profile evolve faster than on a flat surface. 
    %In configuration E, the evolution is qualitatively similar to the flat case, with smaller acceleration in the final stages of the dynamics.
    %The curves for the configurations S$+$ and S$-$ overlap exactly. In both cases, a noticeable deceleration is observed starting around $t\sim13\times10^4$. In configuration P, low velocity is observed during the initial stages.
    }
    \label{fig:sim}
\end{figure}

On such a height profile, we place a dislocation dipole with Burgers vectors $\mathbf{b} = \pm(4\pi/\sqrt{3}, 0)$, as in the previous section. 
The dipole is in a climb configuration, i.e., dislocations are aligned along a direction perpendicular to their Burgers vectors, with initial distance $50 \Lambda_y$. We simulate annihilation by varying the initial relative position of the dipole with respect to the surface bumps, while keeping the initial distance between the dislocations constant. 
To set the same initial condition for all the considered cases, filtering out any impact of the surface profile on the initial relaxation, we first consider a dislocation dipole in a flat system and let it relax for $10^3$ simulation timesteps. 
The resulting amplitudes are used as initial condition for all the simulations on the considered height profiles, shown in Figure~\ref{fig:sim}(a). 
Therein, we refer to each dislocation being positioned at the peak of each bump as P, to dislocations positioned closer to one bump than the other as S$\pm$, and to dislocations positioned in the flat area equidistant from the peaks as E (see labels in Figure~\ref{fig:sim}(a)).    

In all cases, the dislocations attract one another and eventually annihilate. %, meaning that the Peach-Koehler force dominates over the contribution from the height profile. 
Nevertheless, the annihilation process depends on the initial configuration.
In Figure~\ref{fig:sim}(b), we plot the distance between the dislocations in units of the lattice parameter $\Lambda_y$ as a function of time for the different initial configurations. 
When the dislocations are positioned equidistant from the peaks (configuration E), the evolution is qualitatively similar to that in the flat case. A deviation is observed only during the final stages of the dynamics, where the slope of the curve decreases in magnitude, indicating a reduction in dislocation velocity as they approach the bump.
In the configuration S$+$, the top dislocation is initially significantly closer to the bump than the bottom dislocation. As a result, the upper dislocation traverses it prior to annihilation, while the lower dislocation moves over a flat region. 
The interaction with a non-flat region is clearly reflected in the distance plot: around $t\sim13\times10^4$, the slope of the curve decreases in magnitude, corresponding to a reduction in the velocity of the top dislocation as it crosses the bump. 
At $t\sim18\times10^4$, the dislocation clears it, after which its velocity rapidly increases again. 
Notably, the distance plot for the configuration S$-$ is identical to that of S$+$, indicating that dislocations with opposite charge interact in the same way, crossing the bump from below.
Finally, when the dislocations are initially placed at the peaks (configuration P), their velocity remains low up to $t\sim~3000$, corresponding to the dislocations clearing the bumps, after which it increases to match that of other configurations when the dislocations are moving in flat regions. 
The jump at $t=0$ for configuration~P can be attributed to a numerical artifact of the initialization, as the system starts from the amplitudes relaxed in a flat domain and the height profile is imposed instantaneously. This effect is negligible in the other cases.
In short, steep surface profiles significantly affect dislocation velocity. Specifically, the considered dislocations move faster than on a flat surface when far from the bump and slower when near the bump. 
The steepness of the surface profile plays a major role in this behavior. Repeating the study with a pair of gently-varying surface profiles, e.g., defined as in Eq.~\eqref{eq:bumps} with  $a = 2 \Lambda_x$ and $\Delta=10\Lambda_y$, produces dislocation velocities that are indistinguishable from those observed on a flat profile in all configurations considered.

\section{Semi-analytical description of dislocation dynamics} \label{sec:analytica}

The defect velocity \eqref{eq:vv} provides a powerful link between the evolution of the amplitudes $\eta_n$, governed by the underlying free-energy functional, and the dynamics of the topological defects they encode. By means of controlled approximations, we derive a semi-analytical model capturing the impact of non-flat surface profiles on dislocation velocities. Representative cases are explicitly illustrated.

Let us consider one amplitude representing the complex order parameter $\eta=|\eta|e^{i\theta}$ for topological defects. 
To simplify notation, we omit for now the amplitude index $n$. 
The single-valuedness of $\eta$ requires that its magnitude $|\eta|$ vanishes when there is a topological defect in its phase $\arg(\eta)$, i.e. when with $\int_C d\theta = 2\pi s$ with $C$ a close circuit enclosing the defect of charge $s = \pm 1$. 
Hence, topological defects appear as zeros of $\eta$. 
In fact, from the dynamical evolution, $|\eta|$ vanishes linearly inside the defect core corresponding to the stationary solution of $\nabla^2\eta =0$ with the topological constraint at the origin, which implies an inner-core profile given by
\begin{equation}\label{eq:near-core}
    \eta^{\textrm{core}}(\mathbf r) \sim x-isy  \qquad \textrm{ as } \ \ |\mathbf r|\rightarrow 0.    
\end{equation}
This expression for the inner-core amplitude profile leads to the following identities:
\begin{equation}
\begin{split}
\mathrm{Im} \Bigl[\partial_x\eta^{\textrm{core}}\,(\partial_y\eta^{\textrm{core}})^* \Bigr] &= s,
\\[1ex]    
\mathrm{Im}\Bigl[(\partial_x\eta^{\textrm{core}})^*\, \partial_y\eta^{\textrm{core}}\Bigr] &= -s.    
\end{split}
\end{equation}
At the dislocation core $s \neq 0$,  and  from Eq.~\eqref{eq:vv} we get
\begin{equation}\label{eq:v_anl}
\begin{split}
    v_x &= -\frac{1}{s}\mathrm{Im}\Bigl[i s\partial_t\eta^{\textrm{core}} \Bigr]
        = -\,\mathrm{Re}\Bigl[\partial_t\eta^{\textrm{core}} \Bigr],
    \\[1ex]
    v_y &= \frac{1}{s}\mathrm{Im} \Bigl[\partial_t\eta^{\textrm{core}} \Bigr].
\end{split}
\end{equation}
%
% In a complex form $v= v_x+iv_y$, we have that 
% \[v = -\mathrm{Re}[\partial_t\eta^{\textrm{core}}]+i s \mathrm{Im}[\partial_t\eta^{\textrm{core}}] = \frac{s-1}{2}\partial_t\eta^{\textrm{core}}-\frac{s+1}{2}(\partial_t\eta^{\textrm{core}})^*,\]
% where the amplitude evolution of the inner core on a fixed surface is determined by 
From Eq.~\eqref{eq:surf_evo}, the evolution of an amplitude on a fixed surface is given by
\begin{equation}\label{eq:edot_fix}
\partial_t \eta^{\textrm{core}} = -A ||\mathbf{k}||^2_S (\mathcal G^\mathcal{S})^2 \eta^{\textrm{core}},    
\end{equation}
where time is rescaled to normalize the mobility parameter to $M_\eta=1$, and 
\begin{equation}\label{eq:G_eta}
\mathcal G^\mathcal{S} \eta^{\textrm{core}}  = \nabla^2_\mathcal{S} \eta^{\textrm{core}}
+ (\|\mathbf k\|^2-\|\mathbf k_n\|^2_\mathcal{S})\eta^{\textrm{core}}
+ i[2\langle \mathbf k,\nabla_S\eta^{\textrm{core}} \rangle_\mathcal{S}
+ \mathrm{div}_\mathcal{S}(\mathbf k)\eta^{\textrm{core}}]    ,
\end{equation}
with the surface operators as defined in Eq.~\eqref{eq:surf_op}. 
%
%We consider an isolated topological defect in the amplitude $\eta_n$. We first also assume that the physical system is described by only one amplitude (stripe phase \cite{Elder2002,salvalaglio2022coarse}), so that we can omit the amplitude index $n$. We will relax this assumption later.
%For the rest of this section, we omit the amplitude index $n$, with the understanding that all amplitudes have zeros moving with the same velocity. This is necessary to prevent the dislocation from splitting.
%
Evaluating the operator $(\mathcal G^\mathcal{S})^2$ for the inner core solution $\eta^{\textrm{core}}$ from \eqref{eq:near-core} yields
%For small distortions, we neglect gradients of order greater than two:
%
% \begin{equation}
% (\mathcal G^S)^2 \eta \sim 
% % 2\mathcal Q \nabla_S^2 \eta_n 
% -4 \langle \mathbf k,\nabla_S\langle \mathbf k,\nabla_S\eta \rangle_S \rangle_S
% +4i \mathcal Q\langle \mathbf k,\nabla_S\eta \rangle_S ,
% % + \mathcal Q^2 \eta,    
% \end{equation}
%
% where 
% %
% \begin{equation}\label{eq:Q}
%     \mathcal Q = \|\mathbf k\|^2-\|\mathbf k\|^2_S + i \,\mathrm{div}_S(\mathbf k)
% \end{equation} 
% %
% is the zeroth-order coefficient in Eq.~\eqref{eq:G_eta}.
% Notice that $\mathcal \eta = 0$ since we are at the amplitude zero. 
\begin{equation}
(\mathcal G^\mathcal{S})^2 \eta^{\textrm{core}} = 
-4 \langle \mathbf k,\nabla_S\langle \mathbf k,\nabla_S\eta^{\textrm{core}} \rangle_\mathcal{S} \rangle_\mathcal{S}
+4i [\|\mathbf k\|^2-\|\mathbf k\|^2_\mathcal{S} + i \,\mathrm{div}_\mathcal{S}(\mathbf k)] 
    \langle \mathbf k,\nabla_\mathcal{S}\eta^{\textrm{core}} \rangle_\mathcal{S}.
\end{equation}
Due to the linear inner core profile, we get
\begin{equation}
\begin{split}
\langle \mathbf k,\nabla_\mathcal{S}\eta^{\textrm{core}} \rangle_\mathcal{S} &= 
% \mathbf k\cdot\nabla \eta^{\textrm{core}}  = 
(k_{x}-is k_{y}),
\\
\langle \mathbf k,\nabla_\mathcal{S}\langle \mathbf k,\nabla_\mathcal{S}\eta^{\textrm{core}} \rangle_\mathcal{S} \rangle_\mathcal{S} &= 0,  
\end{split}
\end{equation}
and therefore 
\begin{equation}\label{eq:GGeta}
\begin{split}
    (\mathcal G^\mathcal{S})^2 \eta^{\textrm{core}} &=    
4 (s k_y + i k_x) (\|\mathbf k\|^2-\|\mathbf k\|^2_\mathcal{S} + i \,\mathrm{div}_\mathcal{S}(\mathbf k))
\\
&= 4 (s k_y + i k_x)\bigg( \frac{ \langle \nabla h, \mathbf{k} \rangle^2 }{|\mathbf{g}|} 
+ i \frac{ \langle (\nabla \nabla h) \nabla h, \mathbf{k} \rangle }{ |\mathbf{g}| } \bigg)
\\
&\equiv \frac{4}{ |\mathbf{g}|} (s k_y + i k_x) (\mathcal{N}+i\mathcal{D}),
% \nonumber\\
% &=& 4 (s k_y + i k_x) \frac{ 2\langle \nabla h, \mathbf{k} \rangle-\langle \nabla h, \mathbf{k} \rangle^2 +i\langle (\nabla\nabla h)\nabla h, \mathbf{k} \rangle }{|\mathbf{g}|} 
\end{split}
\end{equation}
where in the expansion of $\mathrm{div}_\mathcal{S}(\mathbf{k})$ we used the fact that $\mathbf{k}$ is constant in space, and therefore $\mathrm{div}(\mathbf{k})=0$.
%
% Thus, the defect velocity induced by surface deformation can be written in terms surface operators as  
% \begin{eqnarray*}
%   v &=& -\frac{4A\,\|\mathbf{k}\|_S^2}{|\mathbf g|}
% \left[-s k_y\,\langle\nabla h,\mathbf{k}\rangle 
% +k_x\,\langle(\nabla\nabla h)\nabla h,\mathbf{k}\rangle
% +is\left(k_x\,\langle\nabla h,\mathbf{k}\rangle 
% + s k_y\,\langle(\nabla\nabla h)\nabla h,\mathbf{k}\rangle\right)\right]\nonumber\\
% &\approx& -\frac{4A\,\|\mathbf{k}\|_S^2}{|\mathbf g|}
% \left[-s k_y\,\langle\nabla h,\mathbf{k}\rangle 
% +k_x\,\langle(\nabla\nabla h)\nabla h,\mathbf{k}\rangle
% +is\left(k_x\,\langle\nabla h,\mathbf{k}\rangle 
% + s k_y\,\langle(\nabla\nabla h)\nabla h,\mathbf{k}\rangle\right)\right]
% \end{eqnarray*}
%
%
Then, by combining Eqs.~\eqref{eq:edot_fix} and \eqref{eq:GGeta} into Eq.~\eqref{eq:v_anl}, we obtain analytical expressions for the velocity of the zero of one amplitude moving on a fixed height profile, under the linear core approximation \eqref{eq:near-core}:
\begin{equation}\label{eq:vstripe}
\begin{split}
    v_x &= \frac{4A}{ |\mathbf{g}|} ||\mathbf{k}||^2_\mathcal{S} \,\bigg( s k_y \mathcal{N} - k_x \mathcal{D} \bigg),
    \\
    v_y &= -\frac{4A}{ |\mathbf{g}|}||\mathbf{k}||^2_\mathcal{S} \,\bigg( \frac{1}{s}k_x \mathcal{N} + k_y  \mathcal{D} \bigg).
\end{split}    
\end{equation}
%
% with
% %
% \begin{equation}
% \begin{split}
% % \mathrm{Re}((\mathcal G^S)^2 \eta) &= 
% % \bigg( s k_y \mathcal{N} - k_x \mathcal{D} \bigg) ,
% % \\
% % \mathrm{Im}((\mathcal G^S)^2 \eta) &= 
% % \bigg( k_x \mathcal{N} + s k_y  \mathcal{D} \bigg) ,
% % \\
% \mathcal{N} &= \|\mathbf k\|^2-\|\mathbf k\|^2_\mathcal{S} 
% = \frac{ \langle \nabla h, \mathbf{k} \rangle^2 }{|\mathbf{g}|} ,
% \\
% \mathcal{D} &= \mathrm{div}_\mathcal{S}(\mathbf k) 
% = \frac{ \langle (\nabla \nabla h) \nabla h, \mathbf{k} \rangle }{ |\mathbf{g}| }, 
% \end{split}
% \end{equation}
%
The result above represents the defect velocity in a system described by a single amplitude, such as a stripe phase.
To model systems described by more than one amplitude, we reintroduce the amplitude index $n$, and we define the set of indices of amplitudes that are zero at the core:
\begin{equation}
    \mathcal{I}_0 = \left\{ n \;\middle|\; \eta_n(\mathbf{r}^0) = 0 \right\},    
\end{equation}
using which we can readily extend Eq.~\eqref{eq:vstripe} to

\begin{equation}\label{eq:v_zeros}
\begin{split}
    v_x &= \frac{4A}{ |\mathbf{g}|} \sum_{m\in \mathcal{I}_0}||\mathbf{k}_m||^2_\mathcal{S} \,\bigg( s_m k_m^y \mathcal{N}_m - k_m^x \mathcal{D}_m \bigg),
    \\
    v_y &= -\frac{4A}{ |\mathbf{g}|} \sum_{m\in \mathcal{I}_0} ||\mathbf{k}_m||^2_\mathcal{S} \,\bigg( \frac{1}{s_m}k_m^x \mathcal{N}_m + k_m^y  \mathcal{D}_m \bigg),
    \\
    \mathcal{N}_m &= \langle \nabla h, \mathbf{k}_m \rangle^2  
    =  (k_m^x \partial_x h + k_m^y \partial_y h)^2 \\
    &= (\mathbf k_m \cdot \nabla h)^2,
    \\
    \mathcal{D}_m &= \langle (\nabla \nabla h) \nabla h, \mathbf{k}_m \rangle 
    = k_m^x \partial_x h \partial_{xx} h + k_m^y \partial_y h \partial_{yy} h 
    + (k_m^x \partial_y h + k_m^y \partial_x h) \partial_{xy} h \\
    &= \frac{1}{2}\mathbf k_m\cdot \nabla(|\nabla h|^2),
\end{split}    
\end{equation}
where $k_m^j$ refers to the $j$-th component of vector $\mathbf{k}_m$.
Notice that $\mathcal{N} = \mathcal{O}(|\nabla h|^2)$, while $\mathcal{D} = \mathcal{O}(|\nabla h||\nabla^2 h|)$, 
which causes one contribution to dominate over the other depending on the shape of the height profile considered and, in particular, which of the terms $|\nabla h|$ and $|\nabla^2 h|$ is larger. 

We can further approximate the expression above by expanding $||\mathbf{k}_m||^2_\mathcal{S}$:
% $    \| \mathbf{k}_m \|^2_\mathcal{S} = 
%     \| \mathbf{k}_m \|^2 - { (k_m^x \partial_x h + k_m^y \partial_y h)^2 }/{|\mathbf{g}|} $,
% and noting that the product of the first term     
\begin{equation}\label{eq:vx_approx}
\begin{split}
    v_x &= 
    \frac{4A}{ |\mathbf{g}|} \sum_{m\in \mathcal{I}_0}
    \bigg(\| \mathbf{k}_m \|^2 - {\frac{ \mathcal{N}_m}{|\mathbf{g}|}}\bigg) \,
    \bigg( s_m k_m^y \mathcal{N}_m - k_m^x \mathcal{D}_m \bigg)
    \\
    &=\frac{4A}{ |\mathbf{g}|} \sum_{m\in \mathcal{I}_0} \bigg[
    \| \mathbf{k}_m \|^2 \bigg( s_m k_m^y \mathcal{N}_m - k_m^x \mathcal{D}_m \bigg) 
    - {\frac{ \mathcal{N}_m }{|\mathbf{g}|}} \,
    \bigg( s_m k_m^y \mathcal{N}_m - k_m^x \mathcal{D}_m \bigg)\bigg],
\end{split}    
\end{equation}
and an analogous expression for $v_y$.
Notice that the second term in the sum contains powers of the slope $\partial_j h$ strictly greater than those in the first term. 
Then, if one is interested in regimes in which $|\nabla h|<1$, the second term can be neglected, realizing the approximation $||\mathbf{k}_m||^2_\mathcal{S}\sim||\mathbf{k}_m||^2$ in both components of the velocity.

The expressions above become significantly less cumbersome once a lattice symmetry and a distribution of topological charges have been chosen. 
As an example, we consider the triangular lattice with reciprocal lattice vectors defined in Section~\ref{sec:apfc}, hosting a single dislocation with Burgers vector $\mathbf{b}=(4\pi/\sqrt{3},0)$, resulting in the topological charges  $s_3=-s_1=1$, $s_2=0$.
To lighten the notation, we define $k_x \equiv k_1^x = -k_3^x=-\sqrt{3}/2$, and $k_y\equiv k_1^y= k_3^y=-1/2$, and we note that $\|\mathbf{k}_1\|=\|\mathbf{k}_3\|=1$.   
Then, including the approximation $||\mathbf{k}_m||^2_\mathcal{S}\sim||\mathbf{k}_m||^2$ discussed above, we get
\begin{equation}\label{eq:v_zeros_tri}
\begin{split}    
        v_x &= 
        % \frac{4A}{ |\mathbf{g}|} \bigg[
        % \bigg( s_1 k_1^y \mathcal{N}_1 - k_1^x \mathcal{D}_1 \bigg) +        
        % \bigg( s_3 k_3^y \mathcal{N}_3 - k_3^x \mathcal{D}_3 \bigg)
        % \bigg]
        % \\ &=
        % \frac{4A}{ |\mathbf{g}|} \bigg[
        % -k_y \mathcal{N}_1 - k_x\mathcal{D}_1 + k_y \mathcal{N}_3 + k_x\mathcal{D}_3\bigg]
        % \\ &=
        \frac{4A}{ |\mathbf{g}|} \bigg[
        -2 k_x^2\partial_x h \partial_{xx} h 
        -2 k_x^2 \partial_{y} h \partial_{xy} h  
        -4 {k_x} k_y^2 \partial_x h \partial_{y} h 
        \bigg] ,
         \\ &= 
         \frac{4A}{ |\mathbf{g}|} \bigg[
         -k_x^2\partial_x (|\nabla h|^2)   
         -4 {k_x} k_y^2 \partial_x h \partial_{y} h 
         \bigg],
        \\
        v_y &= 
        % -\frac{4A}{ |\mathbf{g}|} \bigg[
        % \bigg( {s_1}k_1^x \mathcal{N}_1 + k_1^y  \mathcal{D}_1 \bigg) +        
        % \bigg( {s_3}k_3^x \mathcal{N}_3 + k_3^y  \mathcal{D}_3 \bigg)
        % \bigg]
        % \\ &=
        % -\frac{4A}{ |\mathbf{g}|} \bigg[
        % \bigg( - k^x \mathcal{N}_1 + k^y  \mathcal{D}_1 \bigg) +        
        % \bigg( - k^x \mathcal{N}_3 + k^y  \mathcal{D}_3 \bigg)
        % \bigg]
        % \\ &=
        -\frac{4A}{ |\mathbf{g}|} \bigg[
        -2 k_x^3 (\partial_x h)^2 - 2 k_x k_y^2 (\partial_y h)^2  
        +2 k_y^2 \partial_x h \partial_{xy}h + 2 k_y^2 \partial_y h \partial_{yy}h  
        \bigg] .
\end{split}
\end{equation}

% \textcolor{red}{LA: I suggest to split into two equations, one for vx and one vy. Compute them separately. Then we can move this algebra to the appendix and state in the main text the final expression. }
% \textcolor{blue}{MDD: yes, absolutely. I don't intend on reporting the algebra in the final paper, it's just here for now for us to see the steps.}
    
%\textcolor{red}{LA: what happened to the linear leading order term? It looks like now the leading order is non-linear. }
%\textcolor{blue}{MDD: I redid the derivation, and I still get no linear leading order. I believe the issue comes from a typo I had in the definition of the surface norm in Eq.(1.1), now fixed, coming from a possible typo/unusual notation in the same definition in our reference paper.  I used the correct norm for my original calculation, but incorrectly reported the definition of the norm when I wrote eq.(1.1).  The key difference between my result and yours is eq.(6.8). 
%From the corrected eq.(1.1), we get $ \|\mathbf k\|^2-\|\mathbf k\|^2_\mathcal{S} = { \langle \nabla h, \mathbf{k} \rangle^2 }/{|\mathbf{g}|} $, with no terms linear in the slope. To convince myself that the current definition of the norm is correct, I rederived it from scratch in a separate document in this project, surf-norm.tex. }

%\textcolor{red}{LA: Thanks for carefully redoing the calculation. I remember that at some point I forgot to square an expression as you pointed out later, so maybe that's way I obtained the linear term. It seems that the $v_y$ component also have a contribution like $\partial_y|\nabla h|^2$}

\begin{figure}
    \centering
    \includegraphics[width=0.85\linewidth]{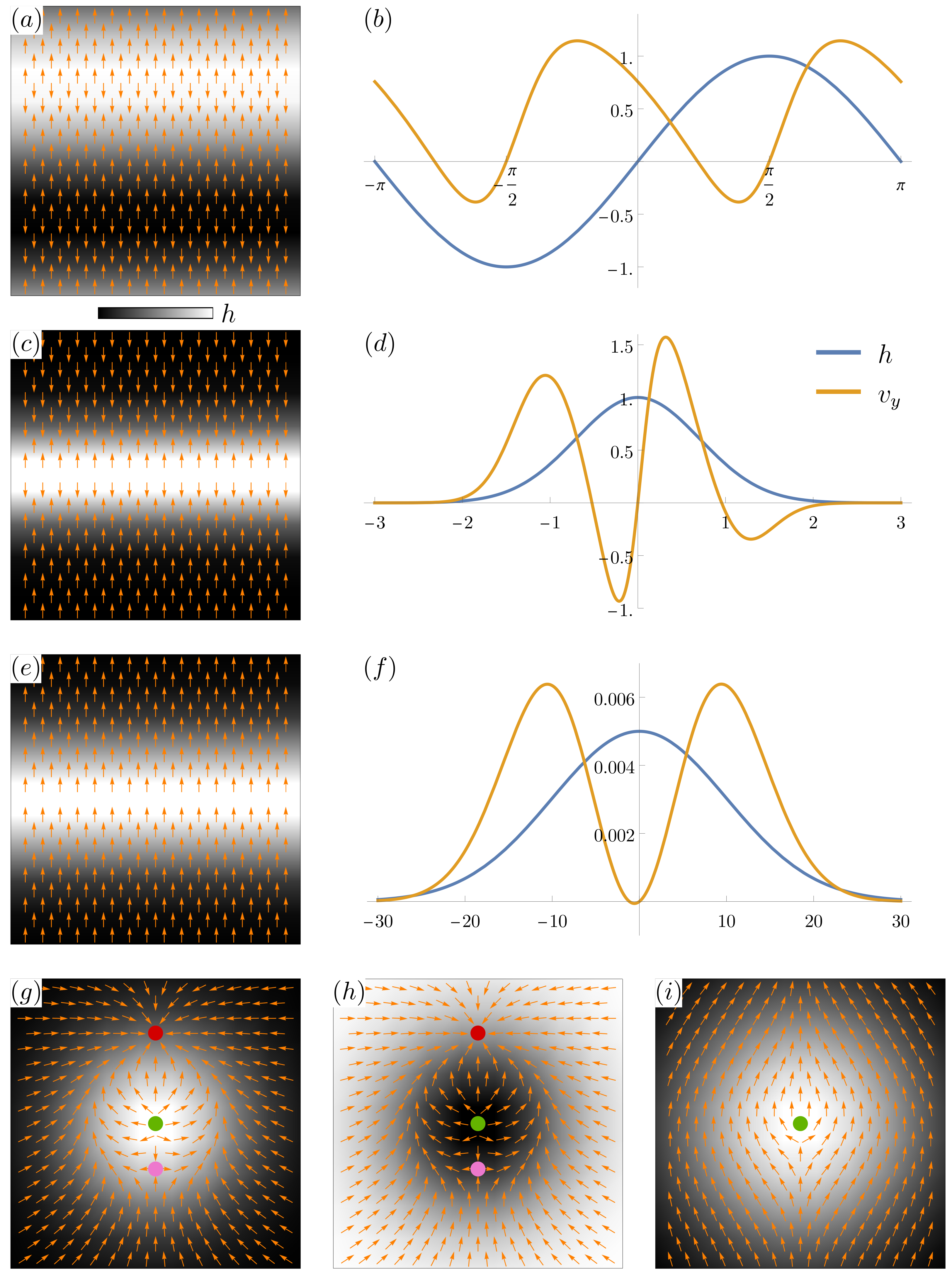}
    \caption{
    Velocity of a single dislocation due to surface deformation in the linear core approximation. The color maps show height profiles, with quivers indicating the direction of dislocation motion.
    One-dimensional plots show the height profile and the $v_y$ component of the velocity across a vertical cut at $x=0$.
    (a,b): Sinusoidal profile. We identify asymmetric stable equilibrium points on the slopes and unstable equilibrium points at the peaks. 
    (c,d): Steep 1D Gaussian. Asymmetric stable equilibrium points on the slopes, one unstable equilibrium point at the peak.  
    (e,f): Shallow 1D Gaussian. The height profile is rescaled by a factor of 200 for plotting purposes. One unstable equilibrium point at the peak. 
    (g,h): Steep 2D Gaussian. Opposite signs of the height profile produce the same velocity field. We identify one stable equilibrium point (red dot), one unstable equilibrium point (green), and one saddle point (pink). The 1D cut is identical to panel (d).
    (i): Shallow 2D Gaussian. Only one unstable equilibrium point (green dot). The 1D cut is identical to panel (f). 
    }
    \label{fig:analytics}
\end{figure}

In Figure~\ref{fig:analytics}, we show plots of the velocity field produced by Eq.~\eqref{eq:v_zeros} for different surface profiles (for the triangular symmetry as considered to obtain \eqref{eq:v_zeros_tri}). %In all cases, we assume a triangular lattice hosting a single dislocation with Burgers vector $\mathbf{b}=(4\pi/\sqrt{3},0)$, resulting in the topological charges $s_3=-s_1=1$, $s_2=0$. 
The dislocation is affected only by the surface geometry, as no external stress and/or other defects are considered. At every point in space, we evaluate the velocity it would acquire if located there.
We show two-dimensional density plots of the height profile and its associated velocity field, along with one-dimensional plots of the vertical cut at $x=0$. 

In Figure~\ref{fig:analytics}(a, b) we impose a one-dimensional sinusoid: $h(x,y) = \sin(y)$. 
We find one unstable equilibrium point at each extremum of the surface and one stable equilibrium on each slope.
Notably, the stable points are not symmetrically distributed along the height profile. In Figure~\ref{fig:analytics}(c, d) we impose a one-dimensional Gaussian: $h(x,y) = e^{-y^2}$. 
We classify this Gaussian as {steep}, as in Eq.~\eqref{eq:v_zeros} the terms $\mathcal{N}_m$ dominate over $\mathcal{D}_m$.
Similarly to the case above, we observe one unstable equilibrium point at the peak of the Gaussian, and two stable equilibrium points on the slopes. Once again, the stable equilibrium points are asymmetric with respect to the height profile. 
In Figure~\ref{fig:analytics}(e, f) we impose a one-dimensional Gaussian: 
$h(x,y) = e^{-y^2/200}$. 
We classify this Gaussian as {shallow}, as in Eq.~\eqref{eq:v_zeros} the curvature terms $\mathcal{D}_m$ dominate.
As expected, the shallower slope of the height profile yields velocity values that are significantly smaller in magnitude than in the previous case. 
However, the overall behavior of the test dislocation moving on this surface is also completely different that the steep profile above. There is only one unstable equilibrium point at the peak of the Gaussian, away from which the dislocation is always pushed in the direction of positive $y$. Note that the velocity contribution due to surface deformation obtained in these cases (Figure~\ref{fig:analytics}(c)-(f)) can be directly related to the results shown in Figure~\ref{fig:sim}, which indicate a transition between accelerated and decelerated motion near and far from the Gaussian bumps, occurring only for steep surface profiles.
In Figure~\ref{fig:analytics}(g, h) we impose a two-dimensional {steep} Gaussian: $h(x,y) = \pm e^{-x^2}e^{-y^2}$. The one-dimensional vertical profile of the surface for $x=0$ is identical to panel Figure~\ref{fig:analytics}(d). 
We use this as an example of a general property of the velocity field we obtain: the field is independent of the sign of the height profile. 
In both cases, we observe exactly one stable equilibrium point (red dot), on the slope of the Gaussian for positive $y$, as well as one unstable equilibrium point on the peak (green dot) and one saddle point (pink dot) on the slope for negative $y$. These semi-analytical results qualitatively reproduce the behavior observed in Figure~\ref{fig:velocity}: dislocations preferentially settle on asymmetric points of the surface slopes. Finally, panel Figure~\ref{fig:analytics}(i) shows a two-dimensional {shallow} Gaussian: 
$h(x,y) = e^{-x^2/200}\,e^{-y^2/200}$. 
The one-dimensional vertical profile of the surface at $x=0$ is identical to that in panel (f).
We identify one unstable equilibrium point at the peak of the Gaussian. Away from it, the dislocation is pushed around the peak and in the positive $y$ direction.

\section{Conclusion}
\label{sec:conclusion}
We have presented a theoretical and computational mesoscale framework for describing the dynamics of dislocations on deformable crystalline surfaces by applying the amplitude-phase-field crystal formalism to curved geometries. This approach resolves crystalline order, defect topology, and surface deformations in a unified manner, bridging the gap between continuum elasticity theories and atomistic models, which are computationally prohibitive for studying curvature-defect coupling over long timescales.

A key contribution of this work is the derivation of a general kinematic expression for dislocation velocity from the APFC evolution equations. This formulation reveals how curvature, bending rigidity, and in-plane stresses combine to generate defect self-propulsion, influence the dislocation glide trajectories, and modify defect-defect interactions. Through numerical simulations, we demonstrated that even modest out-of-plane deformations can substantially influence dislocation dynamics, leading to qualitative departures from the classical behavior expected on flat substrates. These findings underscore the importance of treating surface geometry as a dynamic degree of freedom when modeling defects on deformable surfaces, particularly in soft crystalline systems. Moreover, they suggest that curved surfaces can, in turn, be used to tailor defect dynamics by designing their arrangements and constraining their evolution pathways.

The surface-APFC theoretical framework is thus demonstrated here as a versatile mesoscale modeling tool for studying the fundamental principles governing the coupling of topology, elasticity, and geometry in deformable crystalline surfaces. 
Such complex interplay underlies a broad class of problems in soft and active matter, including curvature-guided defect patterns, defect-mediated morphogenesis in biological tissues, and the mechanics of actively driven elastic sheets, in which the reported evidence and tools might find direct applications. 

\vskip6pt
\enlargethispage{20pt}

\section*{Acknowledgments}

We acknowledge support from the Deutsche Forschungsgemeinschaft (DFG, German Research Foundation), project numbers 447241406 and 417223351 (FOR3013). We also gratefully acknowledge the computing on the high-performance computer at the NHR Center of TU Dresden. This center is jointly supported by the Federal Ministry of Education and Research and the state governments participating in the NHR (\href{www.nhr-verein.de/unsere-partner}{www.nhr-verein.de/unsere-partner}).

%\ack{The authors gratefully acknowledge the Center for Information Services and High-Performance Computing [Zentrum für Informationsdienste und Hochleistungsrechnen (ZIH)] at TU Dresden for providing computing time.}

\appendix

\section{Surface operators and norms}
\label{app:operators}

Differential operators and norms evaluated on the surface $\mathcal{S}\equiv(\mathbf{r},h(\mathbf{r}))$ with $h(\mathbf{r}):\Omega \rightarrow \mathbb{R}$ and $\mathbf{r} \in \Omega \subset  \mathbb{R}^2$ read \cite{Nitschke2020}: %\textcolor{red}{LA: I suggest moving these identities into an Appendix}
\begin{equation}\label{eq:surf_op}
\begin{split}
    \nabla_\mathcal{S} f &= \mathbf{g}^{-1} \cdot \nabla f ,
    \\
    \nabla_\mathcal{S}^2 f &= 
    \nabla^2 f - 
    \frac{ \langle \nabla h, \nabla f \rangle }{ |\mathbf{g}| } \nabla^2 h
    - \frac{ \langle (\nabla h)^2, \nabla\nabla f \rangle }{ |\mathbf{g}| }
    + \frac{ \langle (\nabla h)^2, \nabla \nabla h \rangle 
    \langle \nabla h, \nabla f \rangle} { |\mathbf{g}|^2 } ,
    \\
    \| \mathbf{w} \|^2_\mathcal{S} &= 
    \| \mathbf{w} \|^2 - \frac{ \langle \nabla h, \mathbf{w} \rangle^2 }{|\mathbf{g}|} ,
    \\
    \mathrm{div}_\mathcal{S} (\mathbf{w}) &= 
    \mathrm{div}(\mathbf{w}) + \frac{ \langle (\nabla \nabla h) \nabla h, \mathbf{w} \rangle }{ |\mathbf{g}| },    
    \\
    |\mathbf{g}| &= 1 + (\partial_x h)^2 + (\partial_y h)^2,
    \\
    \langle u,v \rangle_\mathcal{S} &= u^i \mathbf g_{ij}v^j,
\end{split}
\end{equation}
with $f$ a generic function, $\mathbf{w}\in\Omega$ a generic vector, and $\langle \cdot, \cdot \rangle$ the inner product in $\Omega$.


\begin{thebibliography}{29}%
\makeatletter
\providecommand \@ifxundefined [1]{%
 \@ifx{#1\undefined}
}%
\providecommand \@ifnum [1]{%
 \ifnum #1\expandafter \@firstoftwo
 \else \expandafter \@secondoftwo
 \fi
}%
\providecommand \@ifx [1]{%
 \ifx #1\expandafter \@firstoftwo
 \else \expandafter \@secondoftwo
 \fi
}%
\providecommand \natexlab [1]{#1}%
\providecommand \enquote  [1]{``#1''}%
\providecommand \bibnamefont  [1]{#1}%
\providecommand \bibfnamefont [1]{#1}%
\providecommand \citenamefont [1]{#1}%
\providecommand \href@noop [0]{\@secondoftwo}%
\providecommand \href [0]{\begingroup \@sanitize@url \@href}%
\providecommand \@href[1]{\@@startlink{#1}\@@href}%
\providecommand \@@href[1]{\endgroup#1\@@endlink}%
\providecommand \@sanitize@url [0]{\catcode `\\12\catcode `\$12\catcode `\&12\catcode `\#12\catcode `\^12\catcode `\_12\catcode `\%12\relax}%
\providecommand \@@startlink[1]{}%
\providecommand \@@endlink[0]{}%
\providecommand \url  [0]{\begingroup\@sanitize@url \@url }%
\providecommand \@url [1]{\endgroup\@href {#1}{\urlprefix }}%
\providecommand \urlprefix  [0]{URL }%
\providecommand \Eprint [0]{\href }%
\providecommand \doibase [0]{https://doi.org/}%
\providecommand \selectlanguage [0]{\@gobble}%
\providecommand \bibinfo  [0]{\@secondoftwo}%
\providecommand \bibfield  [0]{\@secondoftwo}%
\providecommand \translation [1]{[#1]}%
\providecommand \BibitemOpen [0]{}%
\providecommand \bibitemStop [0]{}%
\providecommand \bibitemNoStop [0]{.\EOS\space}%
\providecommand \EOS [0]{\spacefactor3000\relax}%
\providecommand \BibitemShut  [1]{\csname bibitem#1\endcsname}%
\let\auto@bib@innerbib\@empty
%</preamble>
\bibitem [{\citenamefont {Nelson}(2002)}]{nelson2002defects}%
  \BibitemOpen
  \bibfield  {author} {\bibinfo {author} {\bibfnamefont {D.~R.}\ \bibnamefont {Nelson}},\ }\href@noop {} {\emph {\bibinfo {title} {Defects and geometry in condensed matter physics}}}\ (\bibinfo  {publisher} {\href{https://www.cambridge.org/de/universitypress/subjects/physics/condensed-matter-physics-nanoscience-and-mesoscopic-physics/defects-and-geometry-condensed-matter-physics?format=HB&isbn=9780521801591}{Cambridge University Press}},\ \bibinfo {year} {2002})\BibitemShut {NoStop}%
\bibitem [{\citenamefont {Seung}\ and\ \citenamefont {Nelson}(1988)}]{Seung1988}%
  \BibitemOpen
  \bibfield  {author} {\bibinfo {author} {\bibfnamefont {H.~S.}\ \bibnamefont {Seung}}\ and\ \bibinfo {author} {\bibfnamefont {D.~R.}\ \bibnamefont {Nelson}},\ }\bibfield  {title} {\bibinfo {title} {Defects in flexible membranes with crystalline order},\ }\href {https://doi.org/10.1103/PhysRevA.38.1005} {\bibfield  {journal} {\bibinfo  {journal} {Physical Review A}\ }\textbf {\bibinfo {volume} {38}},\ \bibinfo {pages} {1005} (\bibinfo {year} {1988})}\BibitemShut {NoStop}%
\bibitem [{\citenamefont {Witten}(2007)}]{Witten2007}%
  \BibitemOpen
  \bibfield  {author} {\bibinfo {author} {\bibfnamefont {T.~A.}\ \bibnamefont {Witten}},\ }\bibfield  {title} {\bibinfo {title} {Stress focusing in elastic sheets},\ }\href {https://doi.org/10.1103/RevModPhys.79.643} {\bibfield  {journal} {\bibinfo  {journal} {Reviews of Modern Physics}\ }\textbf {\bibinfo {volume} {79}},\ \bibinfo {pages} {643} (\bibinfo {year} {2007})}\BibitemShut {NoStop}%
\bibitem [{\citenamefont {Guinea}\ \emph {et~al.}(2008)\citenamefont {Guinea}, \citenamefont {Horovitz},\ and\ \citenamefont {Le~Doussal}}]{Guinea2008}%
  \BibitemOpen
  \bibfield  {author} {\bibinfo {author} {\bibfnamefont {F.}~\bibnamefont {Guinea}}, \bibinfo {author} {\bibfnamefont {B.}~\bibnamefont {Horovitz}},\ and\ \bibinfo {author} {\bibfnamefont {P.}~\bibnamefont {Le~Doussal}},\ }\bibfield  {title} {\bibinfo {title} {Gauge field induced by ripples in graphene},\ }\href {https://doi.org/10.1103/PhysRevB.77.205421} {\bibfield  {journal} {\bibinfo  {journal} {Physical Review B}\ }\textbf {\bibinfo {volume} {77}},\ \bibinfo {pages} {205421} (\bibinfo {year} {2008})}\BibitemShut {NoStop}%
\bibitem [{\citenamefont {Zhang}\ \emph {et~al.}(2014)\citenamefont {Zhang}, \citenamefont {Li},\ and\ \citenamefont {Gao}}]{Zhang2014a}%
  \BibitemOpen
  \bibfield  {author} {\bibinfo {author} {\bibfnamefont {T.}~\bibnamefont {Zhang}}, \bibinfo {author} {\bibfnamefont {X.}~\bibnamefont {Li}},\ and\ \bibinfo {author} {\bibfnamefont {H.}~\bibnamefont {Gao}},\ }\bibfield  {title} {\bibinfo {title} {Defects controlled wrinkling and topological design in graphene},\ }\href {https://doi.org/10.1016/j.jmps.2014.02.005} {\bibfield  {journal} {\bibinfo  {journal} {Journal of the Mechanics and Physics of Solids}\ }\textbf {\bibinfo {volume} {67}},\ \bibinfo {pages} {2} (\bibinfo {year} {2014})}\BibitemShut {NoStop}%
\bibitem [{\citenamefont {Roychowdhury}\ and\ \citenamefont {Gupta}(2018)}]{Roychowdhury2018}%
  \BibitemOpen
  \bibfield  {author} {\bibinfo {author} {\bibfnamefont {A.}~\bibnamefont {Roychowdhury}}\ and\ \bibinfo {author} {\bibfnamefont {A.}~\bibnamefont {Gupta}},\ }\bibfield  {title} {\bibinfo {title} {On {{Structured Surfaces}} with {{Defects}}: {{Geometry}}, {{Strain Incompatibility}}, {{Stress Field}}, and {{Natural Shapes}}},\ }\href {https://doi.org/10.1007/s10659-017-9654-1} {\bibfield  {journal} {\bibinfo  {journal} {Journal of Elasticity}\ }\textbf {\bibinfo {volume} {131}},\ \bibinfo {pages} {239} (\bibinfo {year} {2018})}\BibitemShut {NoStop}%
\bibitem [{\citenamefont {Irvine}\ \emph {et~al.}(2010)\citenamefont {Irvine}, \citenamefont {Vitelli},\ and\ \citenamefont {Chaikin}}]{irvine2010pleats}%
  \BibitemOpen
  \bibfield  {author} {\bibinfo {author} {\bibfnamefont {W.~T.}\ \bibnamefont {Irvine}}, \bibinfo {author} {\bibfnamefont {V.}~\bibnamefont {Vitelli}},\ and\ \bibinfo {author} {\bibfnamefont {P.~M.}\ \bibnamefont {Chaikin}},\ }\bibfield  {title} {\bibinfo {title} {Pleats in crystals on curved surfaces},\ }\href {https://doi.org/10.1038/nature09620} {\bibfield  {journal} {\bibinfo  {journal} {Nature}\ }\textbf {\bibinfo {volume} {468}},\ \bibinfo {pages} {947} (\bibinfo {year} {2010})}\BibitemShut {NoStop}%
\bibitem [{\citenamefont {Saw}\ \emph {et~al.}(2017)\citenamefont {Saw}, \citenamefont {Doostmohammadi}, \citenamefont {Nier}, \citenamefont {Kocgozlu}, \citenamefont {Thampi}, \citenamefont {Toyama}, \citenamefont {Marcq}, \citenamefont {Lim}, \citenamefont {Yeomans},\ and\ \citenamefont {Ladoux}}]{saw2017topological}%
  \BibitemOpen
  \bibfield  {author} {\bibinfo {author} {\bibfnamefont {T.~B.}\ \bibnamefont {Saw}}, \bibinfo {author} {\bibfnamefont {A.}~\bibnamefont {Doostmohammadi}}, \bibinfo {author} {\bibfnamefont {V.}~\bibnamefont {Nier}}, \bibinfo {author} {\bibfnamefont {L.}~\bibnamefont {Kocgozlu}}, \bibinfo {author} {\bibfnamefont {S.}~\bibnamefont {Thampi}}, \bibinfo {author} {\bibfnamefont {Y.}~\bibnamefont {Toyama}}, \bibinfo {author} {\bibfnamefont {P.}~\bibnamefont {Marcq}}, \bibinfo {author} {\bibfnamefont {C.~T.}\ \bibnamefont {Lim}}, \bibinfo {author} {\bibfnamefont {J.~M.}\ \bibnamefont {Yeomans}},\ and\ \bibinfo {author} {\bibfnamefont {B.}~\bibnamefont {Ladoux}},\ }\bibfield  {title} {\bibinfo {title} {Topological defects in epithelia govern cell death and extrusion},\ }\href {https://doi.org/10.1038/nature21718} {\bibfield  {journal} {\bibinfo  {journal} {Nature}\ }\textbf {\bibinfo {volume} {544}},\ \bibinfo {pages} {212} (\bibinfo {year} {2017})}\BibitemShut {NoStop}%
\bibitem [{\citenamefont {Angheluta}\ \emph {et~al.}(2025)\citenamefont {Angheluta}, \citenamefont {L{\aa}ng}, \citenamefont {L{\aa}ng},\ and\ \citenamefont {B{\o}e}}]{angheluta2025topological}%
  \BibitemOpen
  \bibfield  {author} {\bibinfo {author} {\bibfnamefont {L.}~\bibnamefont {Angheluta}}, \bibinfo {author} {\bibfnamefont {A.}~\bibnamefont {L{\aa}ng}}, \bibinfo {author} {\bibfnamefont {E.}~\bibnamefont {L{\aa}ng}},\ and\ \bibinfo {author} {\bibfnamefont {S.~O.}\ \bibnamefont {B{\o}e}},\ }\bibfield  {title} {\bibinfo {title} {Topological defects in polar active matter},\ }\href@noop {} {\bibfield  {journal} {\bibinfo  {journal} {\href{https://arxiv.org/abs/2504.03284}{arXiv:2504.03284}}\ } (\bibinfo {year} {2025})}\BibitemShut {NoStop}%
\bibitem [{\citenamefont {Ho}\ \emph {et~al.}(2024)\citenamefont {Ho}, \citenamefont {B{\o}e}, \citenamefont {Dysthe},\ and\ \citenamefont {Angheluta}}]{ho2024role}%
  \BibitemOpen
  \bibfield  {author} {\bibinfo {author} {\bibfnamefont {R.~D.}\ \bibnamefont {Ho}}, \bibinfo {author} {\bibfnamefont {S.~O.}\ \bibnamefont {B{\o}e}}, \bibinfo {author} {\bibfnamefont {D.~K.}\ \bibnamefont {Dysthe}},\ and\ \bibinfo {author} {\bibfnamefont {L.}~\bibnamefont {Angheluta}},\ }\bibfield  {title} {\bibinfo {title} {Role of tissue fluidization and topological defects in epithelial tubulogenesis},\ }\href {https://doi.org/10.1103/PhysRevResearch.6.023315} {\bibfield  {journal} {\bibinfo  {journal} {Physical Review Research}\ }\textbf {\bibinfo {volume} {6}},\ \bibinfo {pages} {023315} (\bibinfo {year} {2024})}\BibitemShut {NoStop}%
\bibitem [{\citenamefont {Bowick}\ and\ \citenamefont {Giomi}(2009)}]{bowick2009two}%
  \BibitemOpen
  \bibfield  {author} {\bibinfo {author} {\bibfnamefont {M.~J.}\ \bibnamefont {Bowick}}\ and\ \bibinfo {author} {\bibfnamefont {L.}~\bibnamefont {Giomi}},\ }\bibfield  {title} {\bibinfo {title} {Two-dimensional matter: order, curvature and defects},\ }\href {https://doi.org/10.1080/00018730903043166} {\bibfield  {journal} {\bibinfo  {journal} {Advances in Physics}\ }\textbf {\bibinfo {volume} {58}},\ \bibinfo {pages} {449} (\bibinfo {year} {2009})}\BibitemShut {NoStop}%
\bibitem [{\citenamefont {Ben~Amar}\ and\ \citenamefont {Pomeau}(1997)}]{Amar1997}%
  \BibitemOpen
  \bibfield  {author} {\bibinfo {author} {\bibfnamefont {M.}~\bibnamefont {Ben~Amar}}\ and\ \bibinfo {author} {\bibfnamefont {Y.}~\bibnamefont {Pomeau}},\ }\bibfield  {title} {\bibinfo {title} {Crumpled paper},\ }\href {https://doi.org/10.1098/rspa.1997.0041} {\bibfield  {journal} {\bibinfo  {journal} {Proceedings of the Royal Society A: Mathematical, Physical and Engineering Sciences}\ }\textbf {\bibinfo {volume} {453}},\ \bibinfo {pages} {729} (\bibinfo {year} {1997})}\BibitemShut {NoStop}%
\bibitem [{\citenamefont {Elder}\ \emph {et~al.}(2002)\citenamefont {Elder}, \citenamefont {Katakowski}, \citenamefont {Haataja},\ and\ \citenamefont {Grant}}]{Elder2002}%
  \BibitemOpen
  \bibfield  {author} {\bibinfo {author} {\bibfnamefont {K.~R.}\ \bibnamefont {Elder}}, \bibinfo {author} {\bibfnamefont {M.}~\bibnamefont {Katakowski}}, \bibinfo {author} {\bibfnamefont {M.}~\bibnamefont {Haataja}},\ and\ \bibinfo {author} {\bibfnamefont {M.}~\bibnamefont {Grant}},\ }\bibfield  {title} {\bibinfo {title} {Modeling {{Elasticity}} in {{Crystal Growth}}},\ }\href {https://doi.org/10.1103/PhysRevLett.88.245701} {\bibfield  {journal} {\bibinfo  {journal} {Physical Review Letters}\ }\textbf {\bibinfo {volume} {88}},\ \bibinfo {pages} {245701} (\bibinfo {year} {2002})}\BibitemShut {NoStop}%
\bibitem [{\citenamefont {Elder}\ and\ \citenamefont {Grant}(2004)}]{Elder2004}%
  \BibitemOpen
  \bibfield  {author} {\bibinfo {author} {\bibfnamefont {K.~R.}\ \bibnamefont {Elder}}\ and\ \bibinfo {author} {\bibfnamefont {M.}~\bibnamefont {Grant}},\ }\bibfield  {title} {\bibinfo {title} {Modeling elastic and plastic deformations in nonequilibrium processing using phase field crystals},\ }\href {https://doi.org/10.1103/PhysRevE.70.051605} {\bibfield  {journal} {\bibinfo  {journal} {Physical Review E}\ }\textbf {\bibinfo {volume} {70}},\ \bibinfo {pages} {051605} (\bibinfo {year} {2004})}\BibitemShut {NoStop}%
\bibitem [{\citenamefont {Emmerich}\ \emph {et~al.}(2012)\citenamefont {Emmerich}, \citenamefont {L{\"o}wen}, \citenamefont {Wittkowski}, \citenamefont {Gruhn}, \citenamefont {T{\'o}th}, \citenamefont {Tegze},\ and\ \citenamefont {Gr{\'a}n{\'a}sy}}]{Emmerich2012}%
  \BibitemOpen
  \bibfield  {author} {\bibinfo {author} {\bibfnamefont {H.}~\bibnamefont {Emmerich}}, \bibinfo {author} {\bibfnamefont {H.}~\bibnamefont {L{\"o}wen}}, \bibinfo {author} {\bibfnamefont {R.}~\bibnamefont {Wittkowski}}, \bibinfo {author} {\bibfnamefont {T.}~\bibnamefont {Gruhn}}, \bibinfo {author} {\bibfnamefont {G.~I.}\ \bibnamefont {T{\'o}th}}, \bibinfo {author} {\bibfnamefont {G.}~\bibnamefont {Tegze}},\ and\ \bibinfo {author} {\bibfnamefont {L.}~\bibnamefont {Gr{\'a}n{\'a}sy}},\ }\bibfield  {title} {\bibinfo {title} {Phase-field-crystal models for condensed matter dynamics on atomic length and diffusive time scales: An overview},\ }\href {https://doi.org/10.1080/00018732.2012.737555} {\bibfield  {journal} {\bibinfo  {journal} {Advances in Physics}\ }\textbf {\bibinfo {volume} {61}},\ \bibinfo {pages} {665} (\bibinfo {year} {2012})}\BibitemShut {NoStop}%
\bibitem [{\citenamefont {Goldenfeld}\ \emph {et~al.}(2005)\citenamefont {Goldenfeld}, \citenamefont {Athreya},\ and\ \citenamefont {Dantzig}}]{Goldenfeld2005}%
  \BibitemOpen
  \bibfield  {author} {\bibinfo {author} {\bibfnamefont {N.}~\bibnamefont {Goldenfeld}}, \bibinfo {author} {\bibfnamefont {B.~P.}\ \bibnamefont {Athreya}},\ and\ \bibinfo {author} {\bibfnamefont {J.~A.}\ \bibnamefont {Dantzig}},\ }\bibfield  {title} {\bibinfo {title} {{Renormalization group approach to multiscale simulation of polycrystalline materials using the phase field crystal model}},\ }\href {https://doi.org/10.1103/PhysRevE.72.020601} {\bibfield  {journal} {\bibinfo  {journal} {Phys. Rev. E}\ }\textbf {\bibinfo {volume} {72}},\ \bibinfo {pages} {020601(R)} (\bibinfo {year} {2005})}\BibitemShut {NoStop}%
\bibitem [{\citenamefont {Salvalaglio}\ and\ \citenamefont {Elder}(2022)}]{salvalaglio2022coarse}%
  \BibitemOpen
  \bibfield  {author} {\bibinfo {author} {\bibfnamefont {M.}~\bibnamefont {Salvalaglio}}\ and\ \bibinfo {author} {\bibfnamefont {K.~R.}\ \bibnamefont {Elder}},\ }\bibfield  {title} {\bibinfo {title} {Coarse-grained modeling of crystals by the amplitude expansion of the phase-field crystal model: an overview},\ }\href {https://doi.org/10.1088/1361-651x/ac681e} {\bibfield  {journal} {\bibinfo  {journal} {Model. Simul. Mater. Sci. Eng.}\ }\textbf {\bibinfo {volume} {30}},\ \bibinfo {pages} {053001} (\bibinfo {year} {2022})}\BibitemShut {NoStop}%
\bibitem [{\citenamefont {Mazenko}(1997)}]{Mazenko97}%
  \BibitemOpen
  \bibfield  {author} {\bibinfo {author} {\bibfnamefont {G.~F.}\ \bibnamefont {Mazenko}},\ }\bibfield  {title} {\bibinfo {title} {Vortex velocities in the $\mathit{O}(\mathit{n})$ symmetric time-dependent ginzburg-landau model},\ }\href {https://doi.org/10.1103/PhysRevLett.78.401} {\bibfield  {journal} {\bibinfo  {journal} {Phys. Rev. Lett.}\ }\textbf {\bibinfo {volume} {78}},\ \bibinfo {pages} {401} (\bibinfo {year} {1997})}\BibitemShut {NoStop}%
\bibitem [{\citenamefont {Skogvoll}\ \emph {et~al.}(2023)\citenamefont {Skogvoll}, \citenamefont {R{\o}nning}, \citenamefont {Salvalaglio},\ and\ \citenamefont {Angheluta}}]{SkogvollNPJ2023}%
  \BibitemOpen
  \bibfield  {author} {\bibinfo {author} {\bibfnamefont {V.}~\bibnamefont {Skogvoll}}, \bibinfo {author} {\bibfnamefont {J.}~\bibnamefont {R{\o}nning}}, \bibinfo {author} {\bibfnamefont {M.}~\bibnamefont {Salvalaglio}},\ and\ \bibinfo {author} {\bibfnamefont {L.}~\bibnamefont {Angheluta}},\ }\bibfield  {title} {\bibinfo {title} {A unified field theory of topological defects and non-linear local excitations},\ }\href {https://doi.org/10.1038/s41524-023-01077-6} {\bibfield  {journal} {\bibinfo  {journal} {npj Comput. Mater.}\ }\textbf {\bibinfo {volume} {9}},\ \bibinfo {pages} {122} (\bibinfo {year} {2023})}\BibitemShut {NoStop}%
\bibitem [{\citenamefont {De~Donno}\ \emph {et~al.}(2024)\citenamefont {De~Donno}, \citenamefont {Angheluta}, \citenamefont {Elder},\ and\ \citenamefont {Salvalaglio}}]{DeDonno2024}%
  \BibitemOpen
  \bibfield  {author} {\bibinfo {author} {\bibfnamefont {M.}~\bibnamefont {De~Donno}}, \bibinfo {author} {\bibfnamefont {L.}~\bibnamefont {Angheluta}}, \bibinfo {author} {\bibfnamefont {K.~R.}\ \bibnamefont {Elder}},\ and\ \bibinfo {author} {\bibfnamefont {M.}~\bibnamefont {Salvalaglio}},\ }\bibfield  {title} {\bibinfo {title} {Mesoscale field theory for quasicrystals},\ }\href {https://doi.org/10.1103/PhysRevResearch.6.043285} {\bibfield  {journal} {\bibinfo  {journal} {Phys. Rev. Res.}\ }\textbf {\bibinfo {volume} {6}},\ \bibinfo {pages} {043285} (\bibinfo {year} {2024})}\BibitemShut {NoStop}%
\bibitem [{\citenamefont {Backofen}\ \emph {et~al.}(2011)\citenamefont {Backofen}, \citenamefont {Gr{\"a}f}, \citenamefont {Potts}, \citenamefont {Praetorius}, \citenamefont {Voigt},\ and\ \citenamefont {Witkowski}}]{Backofen2011}%
  \BibitemOpen
  \bibfield  {author} {\bibinfo {author} {\bibfnamefont {R.}~\bibnamefont {Backofen}}, \bibinfo {author} {\bibfnamefont {M.}~\bibnamefont {Gr{\"a}f}}, \bibinfo {author} {\bibfnamefont {D.}~\bibnamefont {Potts}}, \bibinfo {author} {\bibfnamefont {S.}~\bibnamefont {Praetorius}}, \bibinfo {author} {\bibfnamefont {A.}~\bibnamefont {Voigt}},\ and\ \bibinfo {author} {\bibfnamefont {T.}~\bibnamefont {Witkowski}},\ }\bibfield  {title} {\bibinfo {title} {A {{Continuous Approach}} to {{Discrete Ordering}} on $\mathbb{S}^2$},\ }\href {https://doi.org/10.1137/100787532} {\bibfield  {journal} {\bibinfo  {journal} {Multiscale Modeling \& Simulation}\ }\textbf {\bibinfo {volume} {9}},\ \bibinfo {pages} {314} (\bibinfo {year} {2011})}\BibitemShut {NoStop}%
\bibitem [{\citenamefont {K{\"o}hler}\ \emph {et~al.}(2016)\citenamefont {K{\"o}hler}, \citenamefont {Backofen},\ and\ \citenamefont {Voigt}}]{Kohler2016}%
  \BibitemOpen
  \bibfield  {author} {\bibinfo {author} {\bibfnamefont {C.}~\bibnamefont {K{\"o}hler}}, \bibinfo {author} {\bibfnamefont {R.}~\bibnamefont {Backofen}},\ and\ \bibinfo {author} {\bibfnamefont {A.}~\bibnamefont {Voigt}},\ }\bibfield  {title} {\bibinfo {title} {Stress {{Induced Branching}} of {{Growing Crystals}} on {{Curved Surfaces}}},\ }\href {https://doi.org/10.1103/PhysRevLett.116.135502} {\bibfield  {journal} {\bibinfo  {journal} {Physical Review Letters}\ }\textbf {\bibinfo {volume} {116}},\ \bibinfo {pages} {135502} (\bibinfo {year} {2016})}\BibitemShut {NoStop}%
\bibitem [{\citenamefont {Elder}\ \emph {et~al.}(2021)\citenamefont {Elder}, \citenamefont {Achim}, \citenamefont {Heinonen}, \citenamefont {Granato}, \citenamefont {Ying},\ and\ \citenamefont {{Ala-Nissila}}}]{Elder2021}%
  \BibitemOpen
  \bibfield  {author} {\bibinfo {author} {\bibfnamefont {K.~R.}\ \bibnamefont {Elder}}, \bibinfo {author} {\bibfnamefont {C.~V.}\ \bibnamefont {Achim}}, \bibinfo {author} {\bibfnamefont {V.}~\bibnamefont {Heinonen}}, \bibinfo {author} {\bibfnamefont {E.}~\bibnamefont {Granato}}, \bibinfo {author} {\bibfnamefont {S.~C.}\ \bibnamefont {Ying}},\ and\ \bibinfo {author} {\bibfnamefont {T.}~\bibnamefont {{Ala-Nissila}}},\ }\bibfield  {title} {\bibinfo {title} {Modeling buckling and topological defects in stacked two-dimensional layers of graphene and hexagonal boron nitride},\ }\href {https://doi.org/10.1103/PhysRevMaterials.5.034004} {\bibfield  {journal} {\bibinfo  {journal} {Physical Review Materials}\ }\textbf {\bibinfo {volume} {5}},\ \bibinfo {pages} {034004} (\bibinfo {year} {2021})}\BibitemShut {NoStop}%
\bibitem [{\citenamefont {Benoit–Maréchal}\ \emph {et~al.}(2024)\citenamefont {Benoit–Maréchal}, \citenamefont {Nitschke}, \citenamefont {Voigt},\ and\ \citenamefont {Salvalaglio}}]{Benoit-Marechal_SurfaceAPFC}%
  \BibitemOpen
  \bibfield  {author} {\bibinfo {author} {\bibfnamefont {L.}~\bibnamefont {Benoit–Maréchal}}, \bibinfo {author} {\bibfnamefont {I.}~\bibnamefont {Nitschke}}, \bibinfo {author} {\bibfnamefont {A.}~\bibnamefont {Voigt}},\ and\ \bibinfo {author} {\bibfnamefont {M.}~\bibnamefont {Salvalaglio}},\ }\bibfield  {title} {\bibinfo {title} {Mesoscale modeling of deformations and defects in thin crystalline sheets},\ }\href {https://doi.org/10.1016/j.mechmat.2024.105114} {\bibfield  {journal} {\bibinfo  {journal} {Mechanics of Materials}\ }\textbf {\bibinfo {volume} {198}},\ \bibinfo {pages} {105114} (\bibinfo {year} {2024})}\BibitemShut {NoStop}%
\bibitem [{\citenamefont {Nitschke}\ \emph {et~al.}(2020)\citenamefont {Nitschke}, \citenamefont {Reuther},\ and\ \citenamefont {Voigt}}]{Nitschke2020}%
  \BibitemOpen
  \bibfield  {author} {\bibinfo {author} {\bibfnamefont {I.}~\bibnamefont {Nitschke}}, \bibinfo {author} {\bibfnamefont {S.}~\bibnamefont {Reuther}},\ and\ \bibinfo {author} {\bibfnamefont {A.}~\bibnamefont {Voigt}},\ }\bibfield  {title} {\bibinfo {title} {Liquid crystals on deformable surfaces},\ }\href {https://doi.org/10.1098/rspa.2020.0313} {\bibfield  {journal} {\bibinfo  {journal} {Proceedings of the Royal Society A: Mathematical, Physical and Engineering Sciences}\ }\textbf {\bibinfo {volume} {476}},\ \bibinfo {pages} {20200313} (\bibinfo {year} {2020})}\BibitemShut {NoStop}%
\bibitem [{\citenamefont {Singh}\ \emph {et~al.}(2022)\citenamefont {Singh}, \citenamefont {Pandey},\ and\ \citenamefont {Gupta}}]{singh2022interaction}%
  \BibitemOpen
  \bibfield  {author} {\bibinfo {author} {\bibfnamefont {M.}~\bibnamefont {Singh}}, \bibinfo {author} {\bibfnamefont {A.}~\bibnamefont {Pandey}},\ and\ \bibinfo {author} {\bibfnamefont {A.}~\bibnamefont {Gupta}},\ }\bibfield  {title} {\bibinfo {title} {Interaction of a defect with the reference curvature of an elastic surface},\ }\href {https://doi.org/10.1039/D2SM00126H} {\bibfield  {journal} {\bibinfo  {journal} {Soft Matter}\ }\textbf {\bibinfo {volume} {18}},\ \bibinfo {pages} {2979} (\bibinfo {year} {2022})}\BibitemShut {NoStop}%
\bibitem [{\citenamefont {Helfrich}(1973)}]{Helfrich_bend}%
  \BibitemOpen
  \bibfield  {author} {\bibinfo {author} {\bibfnamefont {W.}~\bibnamefont {Helfrich}},\ }\bibfield  {title} {\bibinfo {title} {Elastic properties of lipid bilayers: Theory and possible experiments},\ }\href {https://doi.org/doi:10.1515/znc-1973-11-1209} {\bibfield  {journal} {\bibinfo  {journal} {Zeitschrift für Naturforschung C}\ }\textbf {\bibinfo {volume} {28}},\ \bibinfo {pages} {693} (\bibinfo {year} {1973})}\BibitemShut {NoStop}%
\bibitem [{\citenamefont {Frigo}\ and\ \citenamefont {Johnson}(2005)}]{FFTW}%
  \BibitemOpen
  \bibfield  {author} {\bibinfo {author} {\bibfnamefont {M.}~\bibnamefont {Frigo}}\ and\ \bibinfo {author} {\bibfnamefont {S.}~\bibnamefont {Johnson}},\ }\bibfield  {title} {\bibinfo {title} {The design and implementation of fftw3},\ }\href {https://doi.org/10.1109/JPROC.2004.840301} {\bibfield  {journal} {\bibinfo  {journal} {Proceedings of the IEEE}\ }\textbf {\bibinfo {volume} {93}},\ \bibinfo {pages} {216} (\bibinfo {year} {2005})}\BibitemShut {NoStop}%
\bibitem [{\citenamefont {Anderson}\ \emph {et~al.}(2017)\citenamefont {Anderson}, \citenamefont {Hirth},\ and\ \citenamefont {Lothe}}]{anderson2017}%
  \BibitemOpen
  \bibfield  {author} {\bibinfo {author} {\bibfnamefont {P.}~\bibnamefont {Anderson}}, \bibinfo {author} {\bibfnamefont {J.}~\bibnamefont {Hirth}},\ and\ \bibinfo {author} {\bibfnamefont {J.}~\bibnamefont {Lothe}},\ }\href@noop {} {\emph {\bibinfo {title} {Theory of Dislocations}}}\ (\bibinfo  {publisher} {\href{http://www.worldcat.org/oclc/1132912878}{Cambridge University Press}},\ \bibinfo {year} {2017})\BibitemShut {NoStop}%
\end{thebibliography}
\end{document}